\documentclass[a4paper,fleqn]{cas-sc}

\usepackage{natbib}
\setcitestyle{authoryear} 
\usepackage{enumitem}
\usepackage{amsmath}
\usepackage[linesnumbered,ruled,vlined]{algorithm2e}
\SetKwInput{KwInput}{Input}                
\SetKwInput{KwOutput}{Output}           

\def\tsc#1{\csdef{#1}{\textsc{\lowercase{#1}}\xspace}}
\tsc{WGM}
\tsc{QE}

\begin{document}
\let\WriteBookmarks\relax
\def\floatpagepagefraction{1}
\def\textpagefraction{.001}

\shorttitle{Processing and Analyzing Real-World Driving Data}    

\shortauthors{J. Han, et al.}  

\title [mode = title]{Processing and Analyzing Real-World Driving Data: Insights on Trips, Scenarios, and Human Driving Behaviors}  

\tnotemark[1] 

\tnotetext[1]{This report and the work described were sponsored by the U.S. Department of Energy (DOE) Vehicle Technologies Office (VTO) under the Systems and Modelling for Accelerated Research in Transportation (SMART) Mobility Laboratory Consortium, an initiative of the Energy Efficient Mobility Systems (EEMS) Program. The following DOE Office of Energy Efficiency and Renewable Energy (EERE) managers played important roles in establishing the project concept, advancing implementation, and providing ongoing guidance: Erin Boyd and Alexis Zubrow.} 

%

\author[1]{Jihun Han}[orcid=0000-0002-3638-2050]

\cormark[1]


\ead{jihun.han@anl.gov}


\credit{Conceptualization, Data curation, Formal analysis, Investigation, Methodology, Software, Validation, Visualization, Writing - original draft, and Writing - review \& editing}

\affiliation[1]{organization={Argonne National Lab},
            city={Lemont},
            postcode={60439}, 
            state={IL},
            country={USA}}

\author[1]{Dominik Karbowski}[]


\ead{dkarbowski@anl.gov}


\credit{Funding acquisition, Project administration, Resources, Supervision, and Writing - review \& editing}

\affiliation[2]{organization={Hyundai-Kia America Technical Center, Inc. (HATCI)},
            city={Superior Charter Township},
            postcode={48198}, 
            state={MI},
            country={USA}}

\author[1]{Ayman Moawad}[]


\ead{amoawad@anl.gov}


\credit{Formal analysis, Methodology, Validation, Visualization, and Writing - review \& editing}

\author[1]{Namdoo Kim}[]


\ead{nkim@anl.gov}


\credit{Methodology, Validation, Visualization, and Writing - review \& editing}

\author[1]{Aymeric Rousseau}[]


\ead{arousseau@anl.gov}


\credit{Funding acquisition, Resources, and Writing - review \& editing}

\author[2]{Shihong Fan}[]


\ead{sfan@hatci.com}


\credit{Data curation and Writing - review \& editing}

\author[2]{Jason Hoon Lee}[]


\ead{hlee@hatci.com}


\credit{Project administration, Resources, Supervision, and Writing - review \& editing}

\author[3]{Jinho Ha}[]


\ead{luckavener@hyundai.com}


\credit{Project administration, Resources, and Writing - review \& editing}

\affiliation[3]{organization={Hyundai Motor Group},
            city={Hwaseong},
            country={South Korea}}

\cortext[1]{Corresponding author}



\begin{abstract}
Analyzing large volumes of real-world driving data is essential for providing meaningful and reliable insights into real-world trips, scenarios, and human driving behaviors. 
To this end, we developed a multi-level data processing approach that adds new information, segments data, and extracts desired parameters. Leveraging a confidential but extensive dataset (over 1 million km), this approach leads to three levels of in-depth analysis: trip, scenario, and driving. 
The trip-level analysis explains representative properties observed in real-world trips, while the scenario-level analysis focuses on scenario conditions resulting from road events that reduce vehicle speed. 
The driving-level analysis identifies the cause of driving regimes for specific situations and characterizes typical human driving behaviors. 
Such analyses can support the design of both trip- and scenario-based tests, the modeling of human drivers, and the establishment of guidelines for connected and automated vehicles.
\end{abstract}


\begin{highlights}
\item Developing and applying multi-level data processing algorithms to two distinct datasets
\item Processing and analyzing large real-world driving data at the trip, scenario, and driving behavior levels
\item Capturing and providing statistically reliable insights into real-world trips, scenarios, and human driving behaviors to support research on verification and validation for connected and automated vehicles.
\end{highlights}


\maketitle

\section{Introduction}\label{sec: intro}

Vehicle systems have unprecedented opportunities to improve their performance thanks to emerging technologies in sensing/perception, driving automation, and vehicle-to-anything (V2X) connectivity (\citealt{vahidi_energy_2018,guanetti_control_2018}). To ensure that connected and automated vehicles (CAVs) can safely and harmoniously drive with surrounding human-driven vehicles on real-world roads, a rigorous verification and validation (V\&V) process is required. 
As the final step, field tests are necessary before introducing CAVs; however, these tests are limited in scale because they are time-consuming and require a significant amount of resources. For these reasons, pure high-fidelity simulations (\citealt{hehenberger_digital_2016}) and anything-in-the-loop (XIL) simulations (a mix of simulations and physical components) (\citealt{schwarz_role_2022}) are considered as a promising approach to accelerate the CAV V\&V process. Specifically, a digital twin approach has been highlighted, as digital twins of physical entities coexist and interact with simulated environments. We also showcased eco-driving for CAVs using various XIL testing setups (\citealt{jeong_-track_2023, han_energy_2023, ard_energy-efficient_2023}).

Collecting and analyzing data is essential for creating realistic simulation scenarios and validating simulation models, enhancing the strength of simulation-based V\&V tests, and leading to successful field tests. To this end, over the past few decades, data has been collected in various ways to build vehicle trajectory datasets. Multiple cameras on poles and drones are used to collect ``bird's-eye'' image data and extract vehicle trajectories through machine learning-based data processing, including object detection and tracking, and cross-camera rectification, e.g., NGSIM (\citealt{us_department_of_transportation_federal_highway_administration_next_2017}), HighD (\citealt{krajewski_highd_2018}), I24-MOTION (\citealt{gloudemans_interstate-24_2023}). On real-world roads, sensor-equipped ego-vehicles are also used to collect perception and motion data during field tests, e.g., Waymo (\citealt{sun_scalability_2020}), NuScene (\citealt{caesar_nuscenes_2020}), Lyft level 5 (\citealt{houston_one_2020}) and pilot studies, e.g., SPMD (\citealt{safety_pilot_model_deployment_safety_2014}). Furthermore, telematics (e.g., Hyundai's BlueLink and GM's OnStar) can be used to collect the CAN-bus data of customer drivers, which represents naturalistic daily trips and driving behaviors of typical drivers, unlike in the pilot studies where drivers are aware of data collection and drive on specified routes. However, such data is not made public due to confidentiality issues.

Data is an important resource for extracting and generating realistic simulation scenarios. These scenarios can be extracted directly from the data or generated randomly by satisfying statistical criteria derived from the data (\citealt{ zhao_accelerated_2017, de_gelder_scenario_2022, yan_learning_2023}). Scenario extraction methods can replicate recorded real-world scenarios, while scenario generation methods can provide a wide range of scenarios, including safety-critical ones (long-tail data), ensuring they remain representative and realistic.
These scenarios are eventually employed to verify and validate the performance of CAVs, including functional safety, driving comforts, and fuel economy. For example, the ISO 26262 standard guideline involves safety-critical tests scenarios to evaluate functional safety (\citealt{riedmaier_survey_2020}). 

While functional safety tests focus on individual scenarios, tests for evaluating fuel economy and driving comfort need to consider representative trips, which consist of a series of scenarios before completing the trip. The current fuel economy tests conducted by the Environmental Protection Agency (EPA) are designed to evaluate a vehicle's fuel economy when it is placed on a chassis dynamometer and follows standard drive cycles that represent real-world trips (e.g., urban and highway) (\citealt{environmental_protection_agency_detailed_nodate}). However, CAVs adapt their driving behaviors based on surrounding road information, indicating that the current tests need improvement due to variations in CAVs' driving patterns (\citealt{mersky_fuel_2016}). Providing data-driven insights into typical trip characteristics is invaluable for generating CAV fuel economy test trips based on road attributes, rather than relying on speed profiles.

Another important aspect of using data is to analyze the driving behaviors of human drivers, which involves understanding how they drive on real-world roads and respond to various road events. Modeling human drivers is a fundamental component of simulations. In microsimulation studies, the data is used to develop car-following models (\citealt{toledo_driving_2007}); however, more focus is the calibration of these models to capture the macroscopic traffic-level metrics (\citealt{kesting_calibrating_2008, menneni_microsimulation_2008}).
The models include both newly proposed ones (\citealt{scanlon_models_2018, han_analytical_2022}) and improved existing models with data-driven parameters (\citealt{hegde_real-world_2021}). Data analysis studies provide valuable insights into driving behaviors to support driver modeling (\citealt{wang_analysis_2019, liu_learning_2022}). 
Furthermore, establishing supervised guidelines for human driving, including defining statistical criteria for acceptable human driving, is crucial for the development of automated driving functions (\citealt{negash_driver_2023}). 

In this paper, we aim to provide valuable and reliable insights into trips, scenarios, and human driving behaviors using the Hyundai customer driving dataset. This dataset is substantial and unique, as it is collected using Hyundai's BlueLink across many customer drivers. It provides a time series of driving data but lacks contextual information that explains why these speed traces change at trip, scenario, and behavior levels. To tackle these challenges, we developed a multi-level data processing approach that is applied to driving data, generating a labeled dataset. Such a way enables us to extract and present statistical information at three different levels, as outlined below.
\begin{itemize}[noitemsep]
    \item Trip-level information: trip characteristics (e.g., traveled distance and frequency of road braking events) and trip composition (e.g., how various scenarios are composed within a single trip).
    \item Scenario-level information: characteristics of braking and cut-in/lane-change scenarios, especially in the context of scenario setup (e.g., approaching speeds, perceivable distances to braking events, and upcoming curvature). 
    \item Behavior-level information: decision-making characteristics (e.g., initiation timing of coasting and braking regimes) and driving behavior characteristics (e.g., acceleration and braking distance, times, and aggressiveness levels, and turning speeds). 
\end{itemize}

The paper is organized as follows: Section 2 briefly introduces a large dataset. Section 3 presents multi-level data processing to prepare data for analysis. Section 4 provides insights derived from data analysis at trip, scenario, and behavior levels. In Section 5, we present our conclusions.

\section{Real-world driving dataset}\label{sec: dataset}

We considered and used the Hyundai customer driving dataset for developing and deploying the data analytics in this paper. 
This dataset was collected by in-vehicle systems with Hyundai's BlueLink services. Our external collaborator, Hyundai America Technical Center Inc. (HATCI), coordinated with Hyundai data teams at the headquarter location and provided the processed Hyundai customer driving data under the collaborative research and development agreements. The Hyundai data team collected customer driving data without personally identifiable information (e.g., rounded geolocation information - only U.S. states identifiable) to ensure confidentiality and stored it on their server. Furthermore, to prevent the potential reveal of the customer identities before data sharing, they opened a virtual server where we accessed and processed the data, which involved segmenting vehicle trajectories for each entire trip into a series of scenario-level trajectories (further details are included in Section \ref{sec:data processing}). With trip-level information extracted from the virtual server, we obtained the Hyundai customer driving dataset\footnote{In this paper, we aim to provide statistical insights through data analysis. Note that we do not present trajectory data, even at the scenario level.} in our encrypted server. This dataset comprises 1,875 vehicles with the following four vehicle models in the U.S., totaling more than 1 million kilometers, with a sample rate of 1 Hz, as summarized in Table \ref{tab:data summary}. 
\begin{itemize}[noitemsep]
    \item SantaFe: Santa Fe 2019 (4-Cyl, GDI, 2.4 Liter)
    \item SantaFe-T: Santa Fe 2019 (4-Cyl, Turbo, GDI, 2.0 Liter)
    \item Sonata: Sonata 2020 (4-Cyl, 2.5 Liter)
    \item Sonata-T: Sonata 2020 (4-Cyl, Turbo, 1.6 Liter)
\end{itemize}
\begin{table}[h]
\caption{Summary of the Hyundai customer driving dataset}
\label{tab:data summary}
\begin{tabular*}{\tblwidth}{@{}L||L|L|L|L@{}}
    \toprule
    {Models} & {\# of Vehicles} & {\# of Trips} & {Distance [km]} & {Time [hr]}  \\ 
    \midrule
    {SantaFe} & {557} & {16,215} & {248,585} & {5,634} \\ 
    {SantaFe-T} & {456} & {11,775} & {178,229} & {3,979} \\ 
    {Sonata} & {78} & {3,474} & {67,624} & {1,349} \\ 
    {Sonata-T} & {784} & {32,053} & {558,464} & {11,940} \\ 
    \midrule
    {Total} & {1,875} & {63,517} & {1,052,902} & {22,902} \\ 
    \bottomrule	
\end{tabular*}
\end{table}

For the purpose of understanding in data processing in the Section \ref{sec:data processing}, we also considered another dataset, the safety pilot model deployment (SPMD) dataset. This dataset was collected by vehicles equipped with data acquisition systems in the SPMD program sponsored by the U.S. Department of Transportation. The SPMD dataset is publicly available on the Intelligent Transportation Systems (ITS) DataHub\footnote{\url{https://its.dot.gov/data/}}. We used the driving dataset provided by University of Michigan Transportation Research Institute (UMTRI). The SPMD UMTRI dataset provides real-world driving data of more than 110,000 miles of 99 vehicles in Ann Arbor,  with a sample rate of 10 Hz, including geolocation information. All plots shown in the Section \ref{sec:data processing} are based on the SPMD UMTRI datset. 

\begin{figure}[h!]
    \centering    \includegraphics[width=0.6\columnwidth,keepaspectratio]{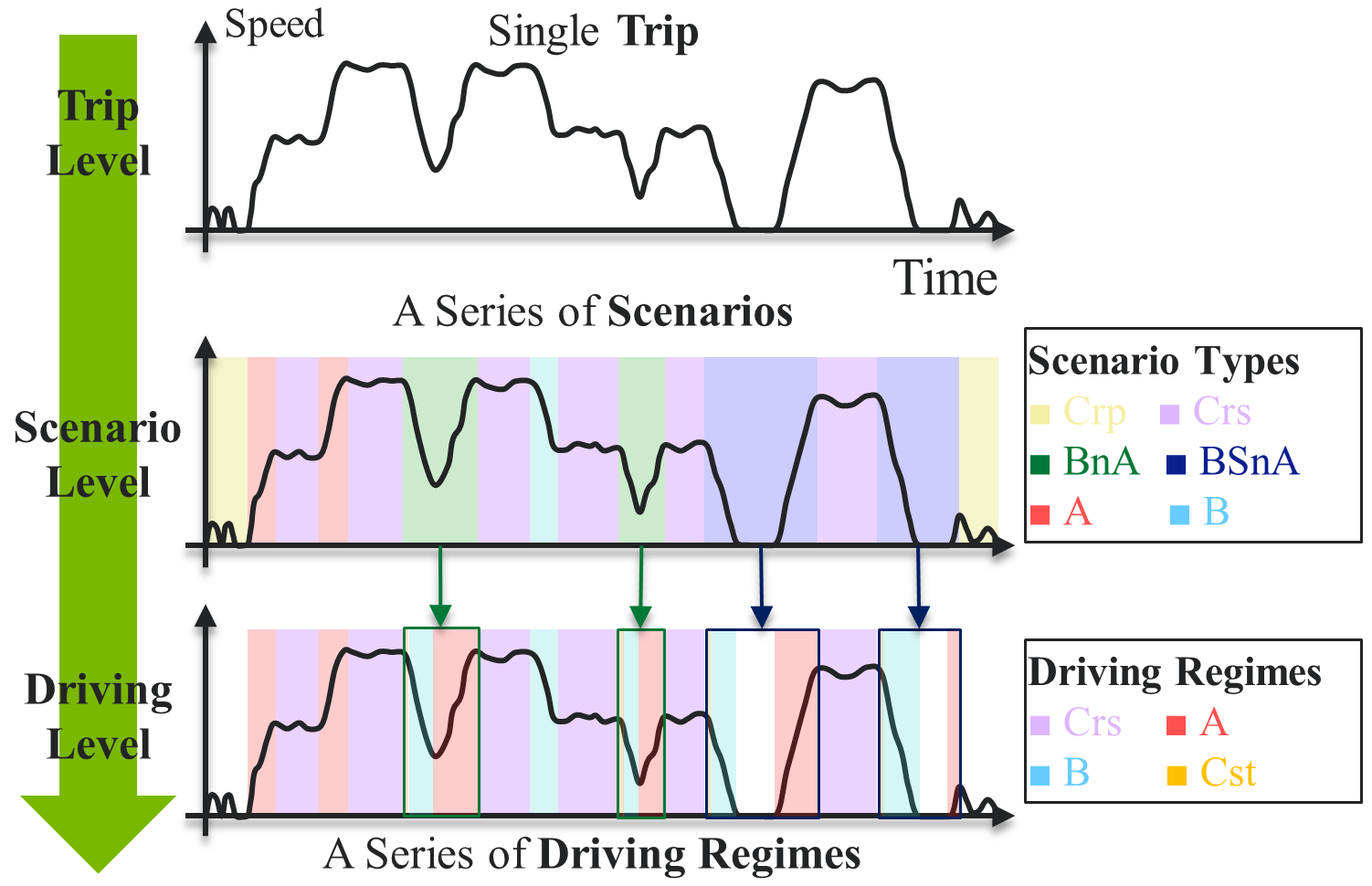}
    \caption{A schematic diagram for multi-level data processing. Note that the abbreviations in scenario types and driving regimes are: creeping (Crp), cruising (Crs), braking and acceleration (BnA), braking, stop, and acceleration (BSnA), acceleration (A), braking (B), and coasting (Cst). Definitions of these abbreviations are included in the following subsections.}
    \label{fig:data proc}
\end{figure}

\section{Multi-level data processing}\label{sec:data processing}

As shown in Figure \ref{fig:data proc}, the proposed data processing consists of three levels depending on the focus: 1) trip level, 2) scenario level, and 3) driving behavior level. The processing for each level is explained in the following sections.

\subsection{Trip-level process: trip parameter extraction}

\begin{figure}[h!]
    \centering    \includegraphics[width=0.98\columnwidth,keepaspectratio]{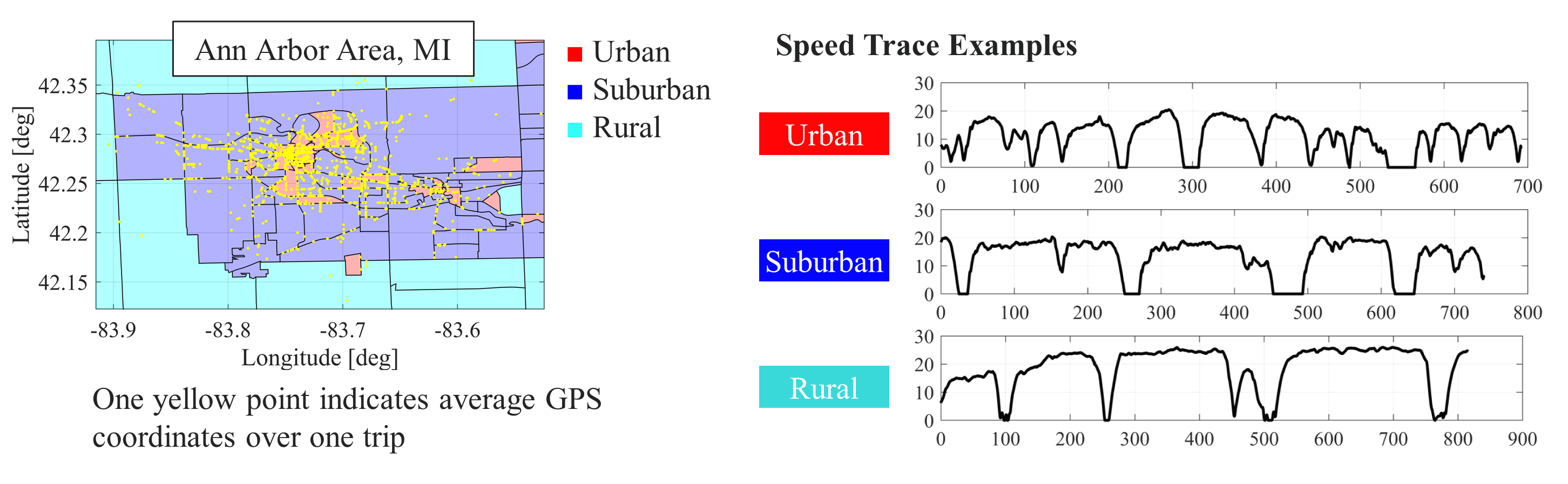}
    \caption{UPSAI-based neighborhood types (left) in Ann Arbor area, MI and speed trace data (right) for each trip type using the SPMD dataset.}
    \label{fig:trip proc}
\end{figure}

As the dataset provides a time series of driving data for an entire trip, the trip-level process extracts trip parameters (e.g., traveled distance, average speed, fuel economy) and provides high-level information. Moreover, this level of processing can categorize each trip's data into different types (e.g., urban, suburban, and highway trips) to capture the differences in driving on various road types. To identify trip types, the Urbanization Perceptions Small Area Index (UPSAI) can be used (\citealt{bucholtz_urbanization_2020}). The UPSAI is a machine learning model developed using neighborhood description data from the 2017 American Housing Survey combined with U.S. census tract boundary data. The UPSAI can predict the likelihood of the neighborhood types for a given area (e.g., urban, suburban, and rural). If the driving data contains accurate GPS coordinates, we can easily deploy the UPSAI model to identify the trip type by simply checking which tract-level area (formed by polygons) includes the recorded GPS coordinate data points. As shown in Figure \ref{fig:trip proc}, the trip type can be identified as a suburban trip if GPS coordinates (yellow points) are within blue areas. On the other hand, if the driving data lacks GPS coordinates or contains inaccurate ones (e.g., rounded-off values due to privacy), a clustering methodology trained on GPS-contained data can be used to identify the trip type using the other signals. For example, the average speed\footnote{Average speed can be computed by dividing the traveled distance by the travel time for each trip.} can be an influential factor in determining trip type, as it correlates with road types. Note that suburban road driving will have higher average speeds than urban road driving due to higher maximum speed limits and longer distances between intersections.

\subsection{Scenario-level process: scenario segmentation}\label{subsec: scen lvl}
As each trip involves various scenarios, the scenario-level process segments the trip data into predefined scenario types.
We firstly defined six scenario types based on the significant speed changes resulting from upcoming road events as follows:  
\begin{itemize}[noitemsep]
    \item Cruising (Crs): maintain a desired constant speed with small variation.
    \item Braking, stop, and acceleration (BSnA): decelerate to stop, wait for a while, and then accelerate, due to traffic lights, stop signs, etc.
    \item Braking and acceleration (BnA): decelerate and then accelerate without a stop, due to traffic lights, curved roads/roundabouts/speed bumps, etc.
    \item Acceleration (A): accelerate due to the increase in the maximum speed limit.
    \item Braking (B): decelerate due to the decrease in the maximum speed limit.
    \item Creeping (Crp): move forward at a very low speed including stops (e.g., searching a parking spot).    
\end{itemize}
All scenarios can have different conditions by depending on the presence of the preceding vehicle (i.e., free-flow vs. car-following) or the involvement of turning (driving straight vs. driving with turns). 

\begin{figure}[h!]
    \centering  
    \includegraphics[width=0.9\columnwidth,keepaspectratio]{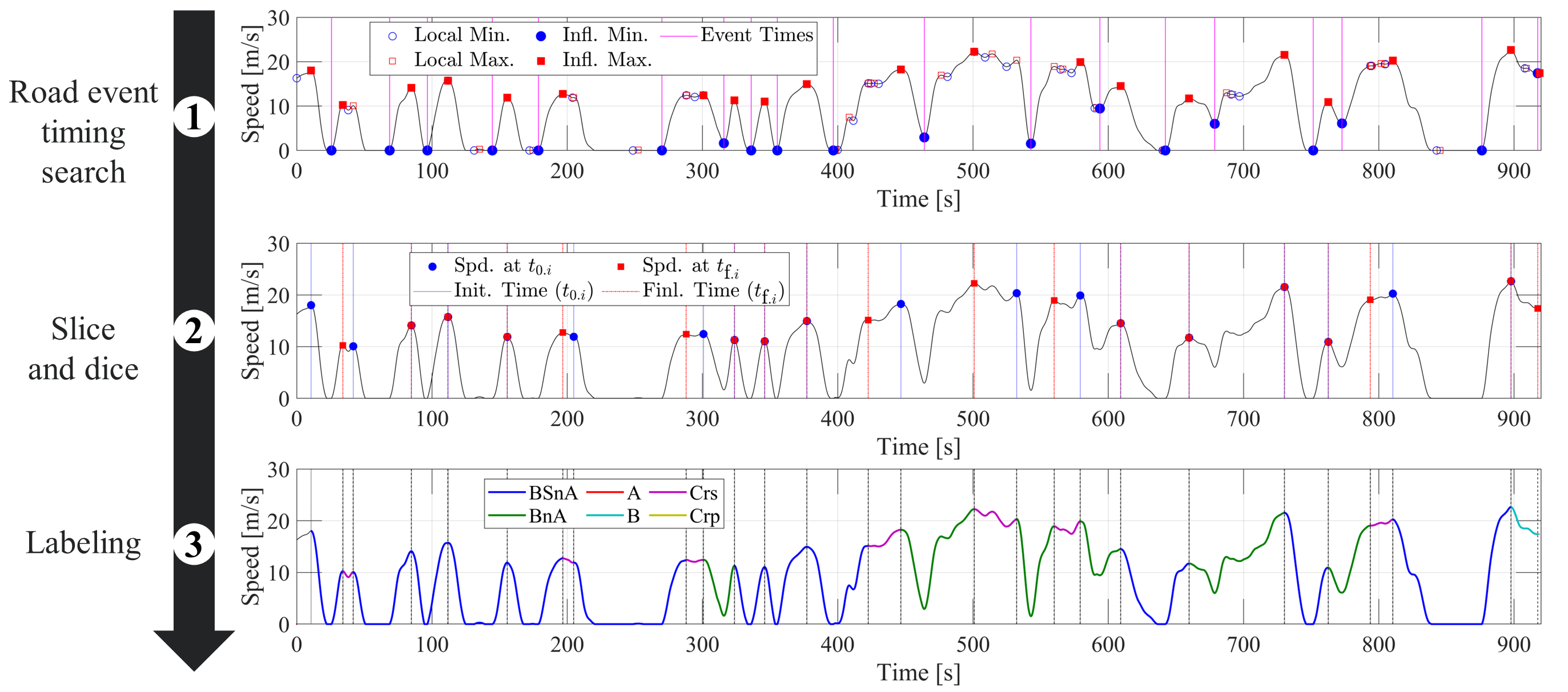}
    \caption{Schematic of the scenario segmentation algorithm with one trip data from the SPMD dataset.}
    \label{fig:scen proc}
\end{figure}

A segmentation algorithm segments a time series of driving data and classifies the segmented data into six scenario types, as shown in Figure \ref{fig:scen proc}. There are three main steps:  
\begin{enumerate}[noitemsep, label=\arabic*)]
    \item Road event timing search: find local extremes (i.e., maximums and minimums) and select influential local extremes among them.    
    \item Slice and dice: define time intervals based on the selected influential local extremes and break entire trip data into smaller parts.    
    \item Labeling: classify each scenario-level segmented data into predefined scenario types.   
\end{enumerate}
Figure \ref{fig:scen proc} shows an example of single trip data where numerous events cause braking, followed by acceleration, and possible stops. Using algorithms we developed (details are included in Algorithm \ref{alg1} and Algorithm \ref{alg2}\footnote{Crs and Crp scenarios are automatically segmented based on the time interval between two consecutive scenarios; any cases including at least one stop are considered as the Crp scenario.} in Appendix \ref{App: codes}), the entire trip data can be segmented, and each scenario can be categorized. For example, segmented driving data is classified as the BSnA scenario type if it consists of braking, stop, and acceleration within its time interval. Note that the threshold for speed increase and decrease to determine whether local extremes are influential can be adjusted, although we have set this threshold to 5 m/s.

\subsection{Driving-level process: driving regime isolation}
\begin{figure}[h!]
    \centering
    \includegraphics[width=0.3\columnwidth,keepaspectratio]{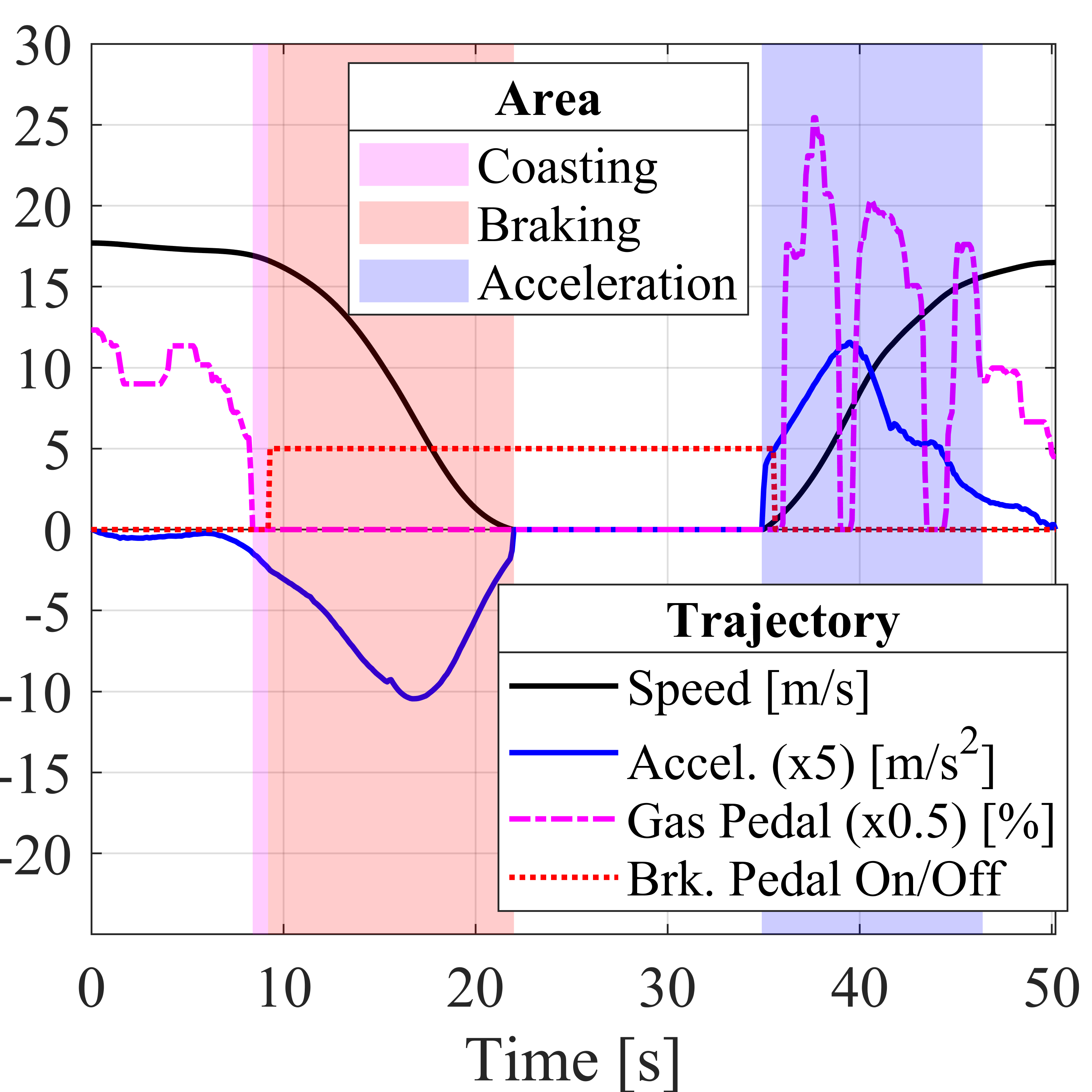} \qquad \qquad
    \includegraphics[width=0.3\columnwidth,keepaspectratio]{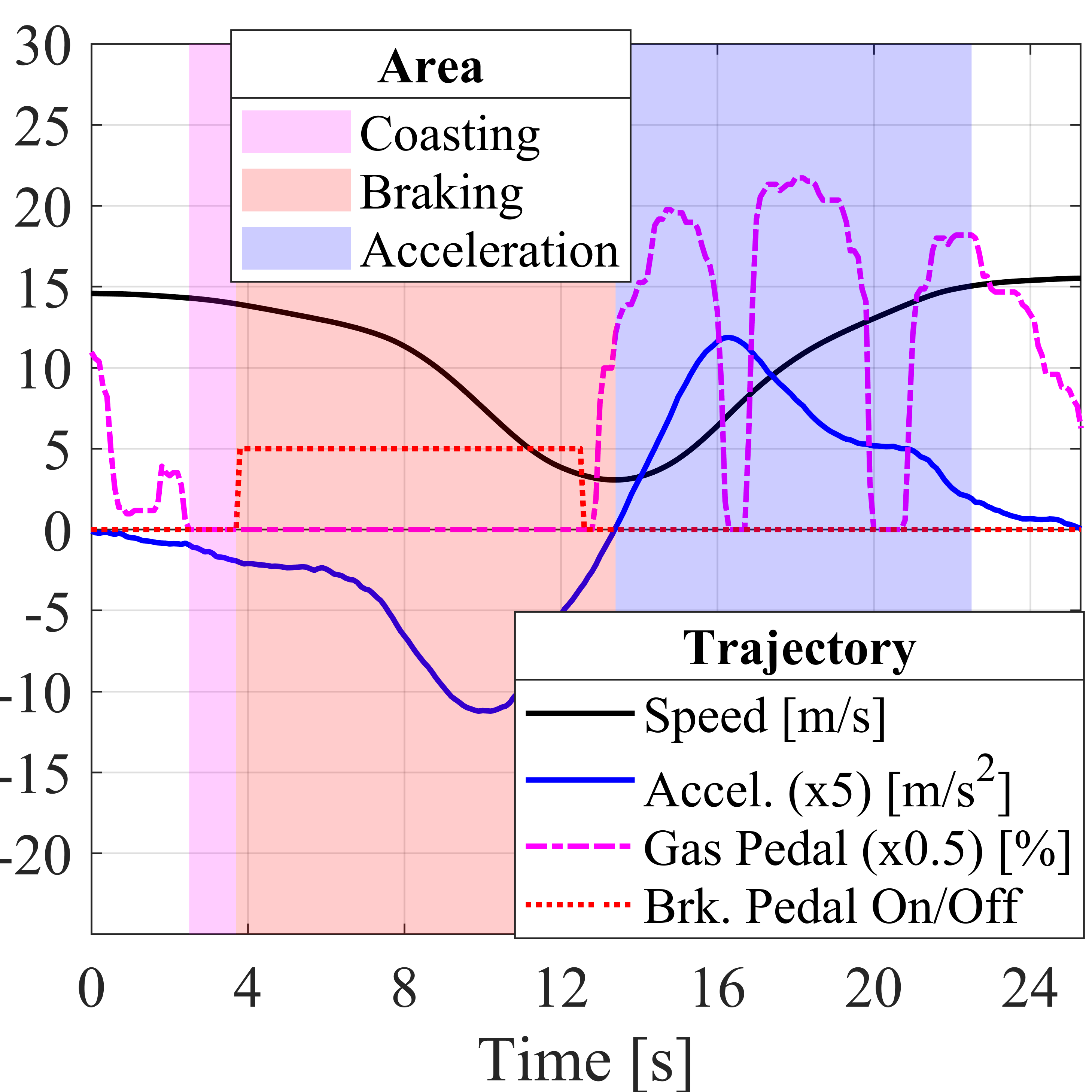}
    \caption{Driving regime isolation result applying to specific BSnA (left) and BnA (right) scenario data examples from the SPMD dataset. The brake pedal values of 5 and 0 indicate whether the brake pedal is pressed (on) or not (off), respectively.}
    \label{fig:drv proc}
\end{figure}

The driving-level process further divides each scenario and isolates driving regimes including the coasting, braking, and acceleration regimes, as shown in Figure \ref{fig:drv proc}. The definition of each driving regime is detailed below. 
\begin{itemize}[noitemsep]
    \item The coasting (Cst) regime: the interval from the moment that the pedals are not depressed to the moment that the brake pedal is depressed.
    \item The braking regime (B): the interval from the moment that the brake pedal is depressed and its pressure increases to the moment that the desired minimum speed is reached. 
    \item The accelerating regime (A): the interval from the moment that the gas pedal is depressed and its pressure increases to the moment that the acceleration is below its threshold (e.g., 0.2 m/s$^2$) after the speed enters within an allowable range that is determined by the final speed using margins (±2 m/s).  
\end{itemize}
Note that we established the rules by the above definition for each driving regime, and applied the driving regime isolation algorithm to scenario-level data. 
Figure \ref{fig:drv proc} shows real-world human driving behavior examples for two BSnA and BnA scenarios. After a driver perceived upcoming road event (e.g., red light at the intersection), he/she released the both pedals and slowed down the speed. Then, the driver started pressing the brake pedal to increase the deceleration, followed by reducing the deceleration enabling the smooth transition to either stopping or acceleration. After the road event is changed (e.g., green light back on), he/she started pressing the gas pedal to increase the acceleration and then adjusted its pressure to gradually decrease the acceleration until the desired speed is reached. 

\section{Exploratory data analysis}\label{sec: data analysis}

\subsection{Trip-level insights}\label{subsec:trip}
\begin{figure}[h!]
    \centering
    \includegraphics[width=0.5\columnwidth,keepaspectratio]{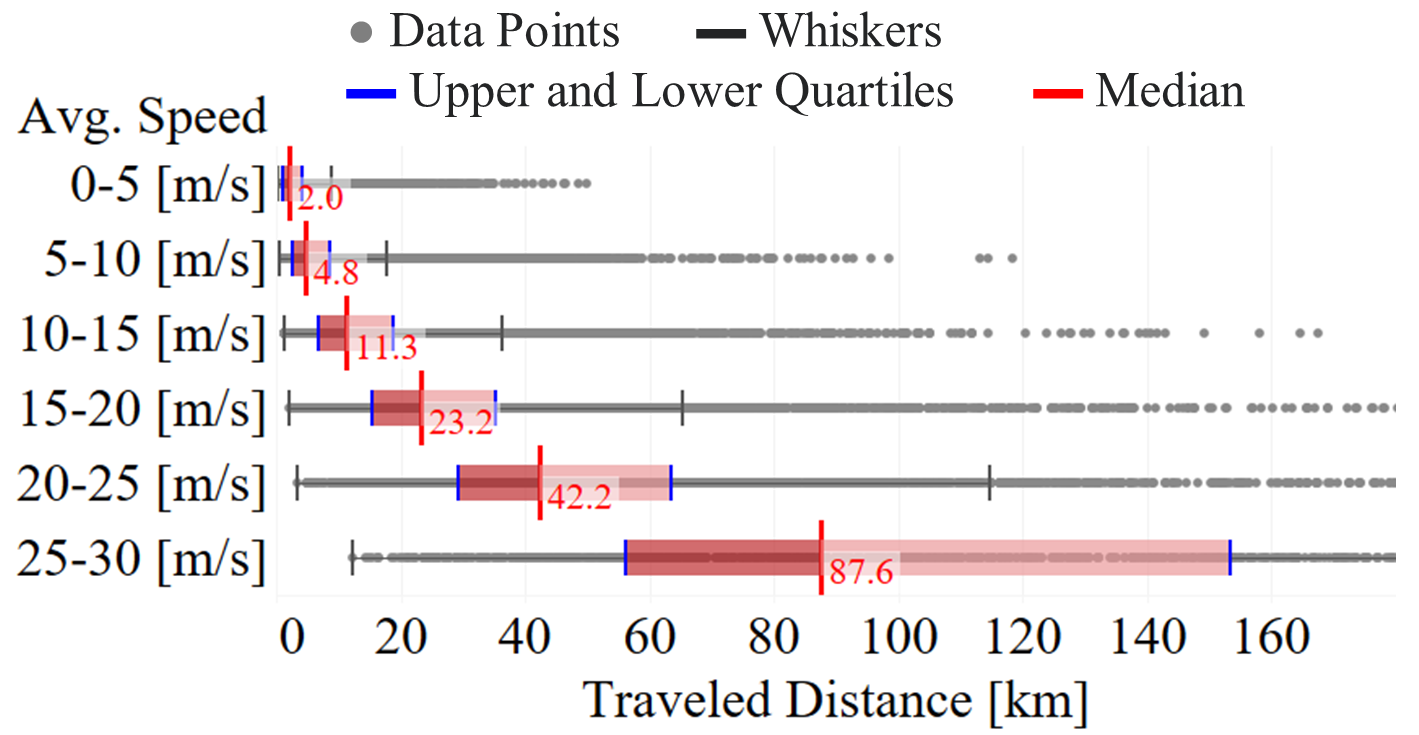} 
    \caption{Traveled distance depending on the average speed range.}
    \label{fig:trip distance}
\end{figure}

\begin{figure}[h!]
    \centering
    \includegraphics[width=0.5\columnwidth,keepaspectratio]{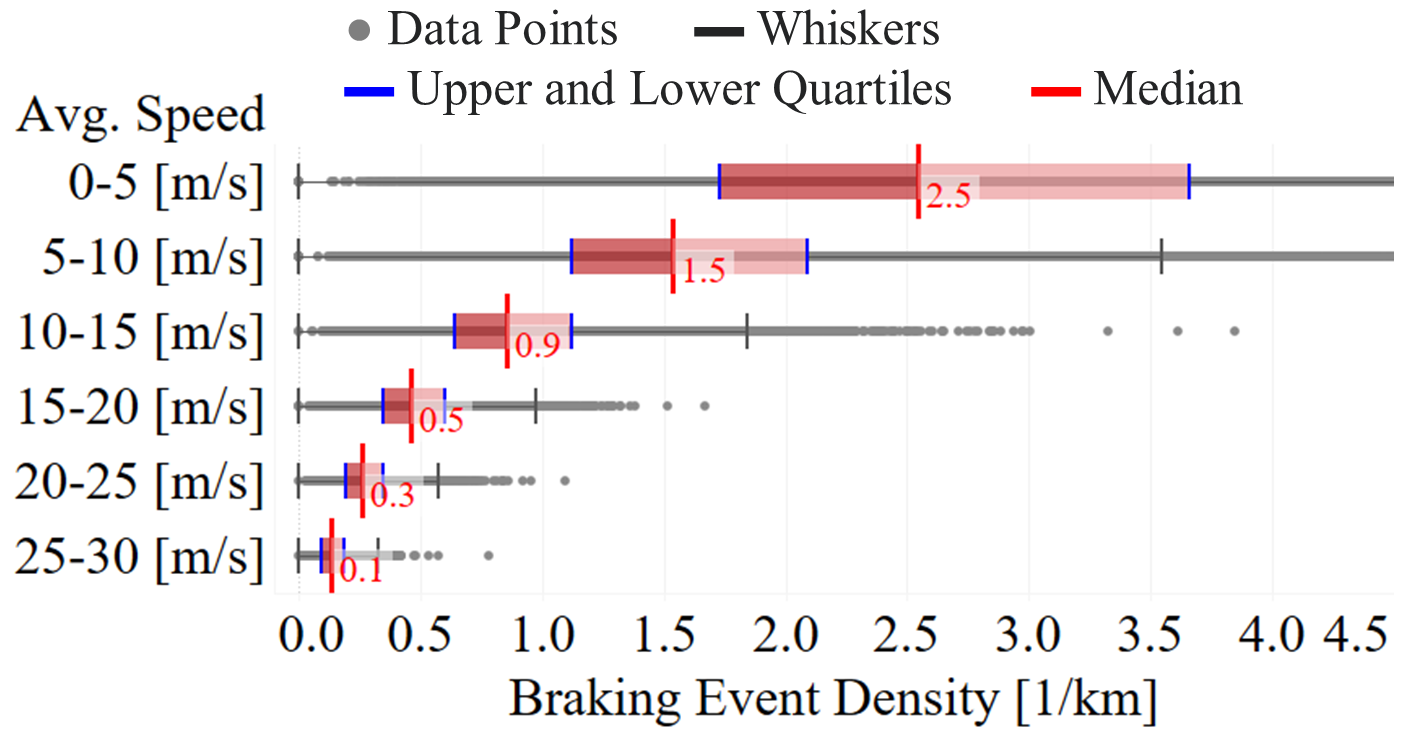} 
    \caption{Braking event density depending on the average speed range.}
    \label{fig:trip braking event density}
\end{figure}

\begin{figure}[h!]
    \centering
    \includegraphics[width=0.5\columnwidth,keepaspectratio]{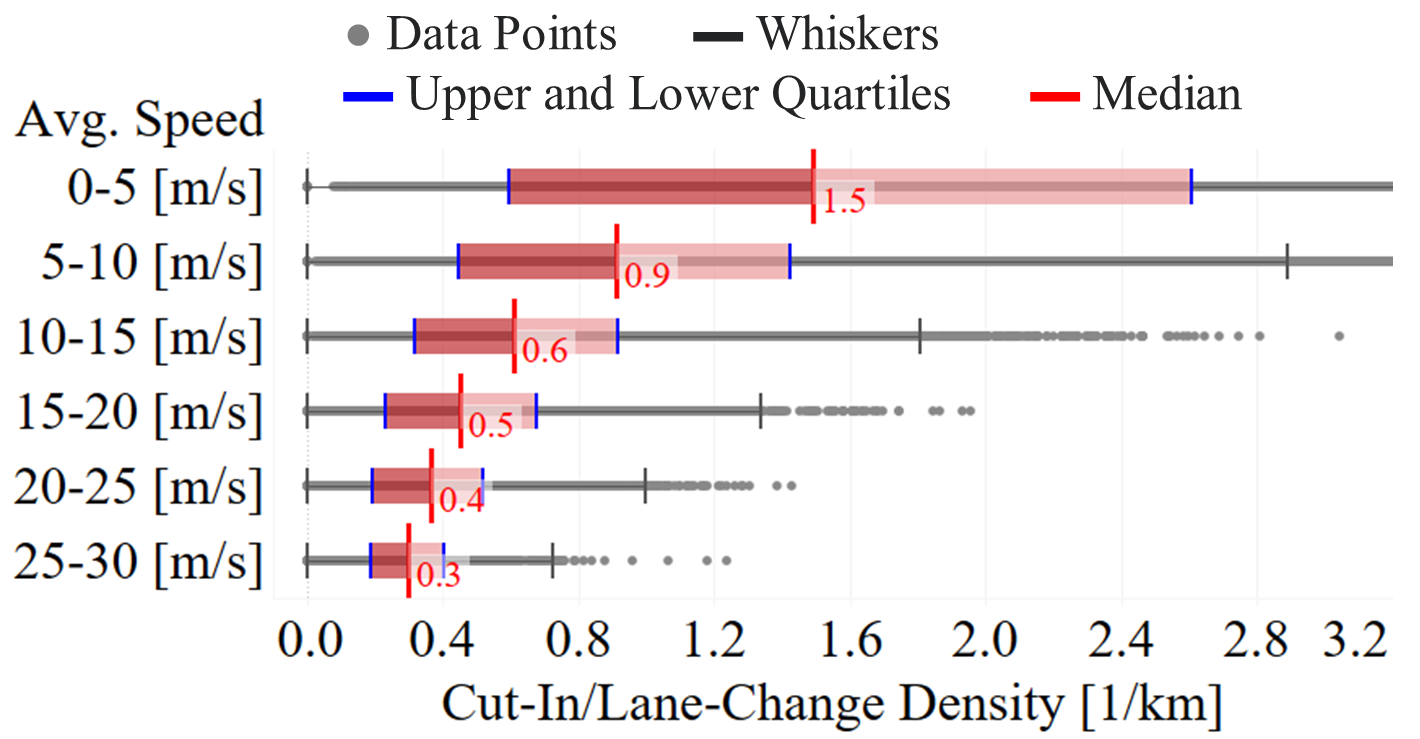} 
    \caption{Cut-in/lane-change density depending on the average speed range.}
    \label{fig:trip cutin density}
\end{figure}

Analyzing customer driving data allows us to investigate the daily trips of Hyundai customers for specific purposes, such as commuting or grocery shopping. These insights are useful for designing single trip test scenarios to validate CAVs. Trip test scenarios should be realistic based on the data-driven insights; otherwise, CAV impacts (e.g., energy savings) may be overestimated or underestimated.  
In the HATCI dataset, individual trips are distinguished by using the ignition signal, with a single trip defined as the period between ignition being turned on (the starting point) and off (the destination). Three trip parameters are considered to represent important trip characteristics, as follows:
\begin{itemize}[noitemsep]
    \item Traveled distance for each trip. 
    \item Braking event density: the number of occurrences of braking road events (i.e., BSnA and BnA situations) per kilometer.   
    \item Cut-in/lane-change\footnote{Cut-in and lane-change events can be identified by checking the time instances at which the distance gap suddenly decreases. Such a sudden decrease is defined as a decrease in the distance gap that exceeds what would be expected based on the relative speed (i.e., speed difference with the preceding vehicle), with an added margin. For example, we consider it a cut-in/lane-change event if the distance gap drops to 20 m from 60 m with a relative speed of -5 m/s, since a distance gap decrease of 40 m cannot occur within the sampling time of 1 s.} density: the number of occurrences where the preceding vehicle suddenly appears in front per kilometer.  
\end{itemize}
Figures \ref{fig:trip distance}, \ref{fig:trip braking event density}, and \ref{fig:trip cutin density} show statistical summaries of these three parameters including median, quartiles, and whiskers. 
Trips that involve rural or highway driving, where higher maximum speeds are permitted, tend to be higher average speeds. As shown in Figure \ref{fig:trip distance}, higher average speeds become longer trips with longer traveled distances. For example, typical highway trips (average speed of 25--30 m/s) range between 50 km and 160 km, with a medium value of 87.6 km. With lower average speeds, the distance range between lower and upper quartiles becomes narrower. This is because of the lower standard deviation of traveled times.   

As shown in Figure \ref{fig:trip braking event density}, the increase in braking event density at lower average speed aligns with the understanding that urban or city trips involve corridors where drivers more often brake in response to road signs, such as red traffic lights. 
Additionally, the distance between two intersections becomes shorter compared to suburban trips. 
In highway trips, traffic congestion is the primary reason for speed reduction, but it does not require as much braking as needed to be considered a braking event, resulting in fewer braking events. 

Figure \ref{fig:trip cutin density} shows that the density of preceding vehicle appearances decreases with higher average speeds. This trend is consistent with the fact that drivers tend to be more cautious during higher-speed driving when changing lanes. Consequently, they require more space between the preceding vehicle and the following vehicle in the target lane, resulting in a lower density of preceding vehicle appearances.
For urban trips with an average speed of 5-10 m/s over the medium distance of 4.8 km, drivers commonly change lanes or experience cut-ins about 4 times, calculated by the medium density of 0.9.

\begin{figure}[h!]
    \centering    \includegraphics[width=0.5\columnwidth,keepaspectratio]{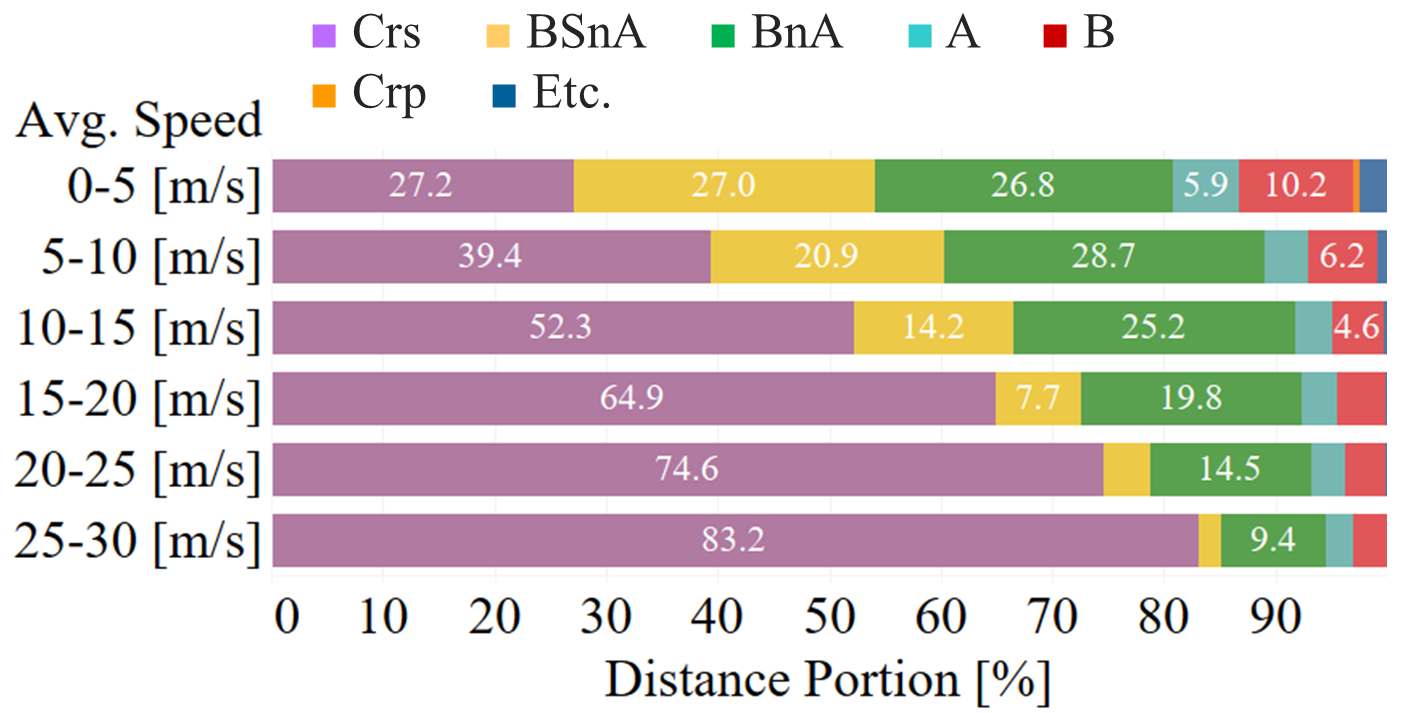} 
    \caption{Scenario type distribution depending on the average speed range.}
    \label{fig:trip scenario distribution}
\end{figure}

\begin{figure}[h!]
    \centering    \includegraphics[width=0.5\columnwidth,keepaspectratio]{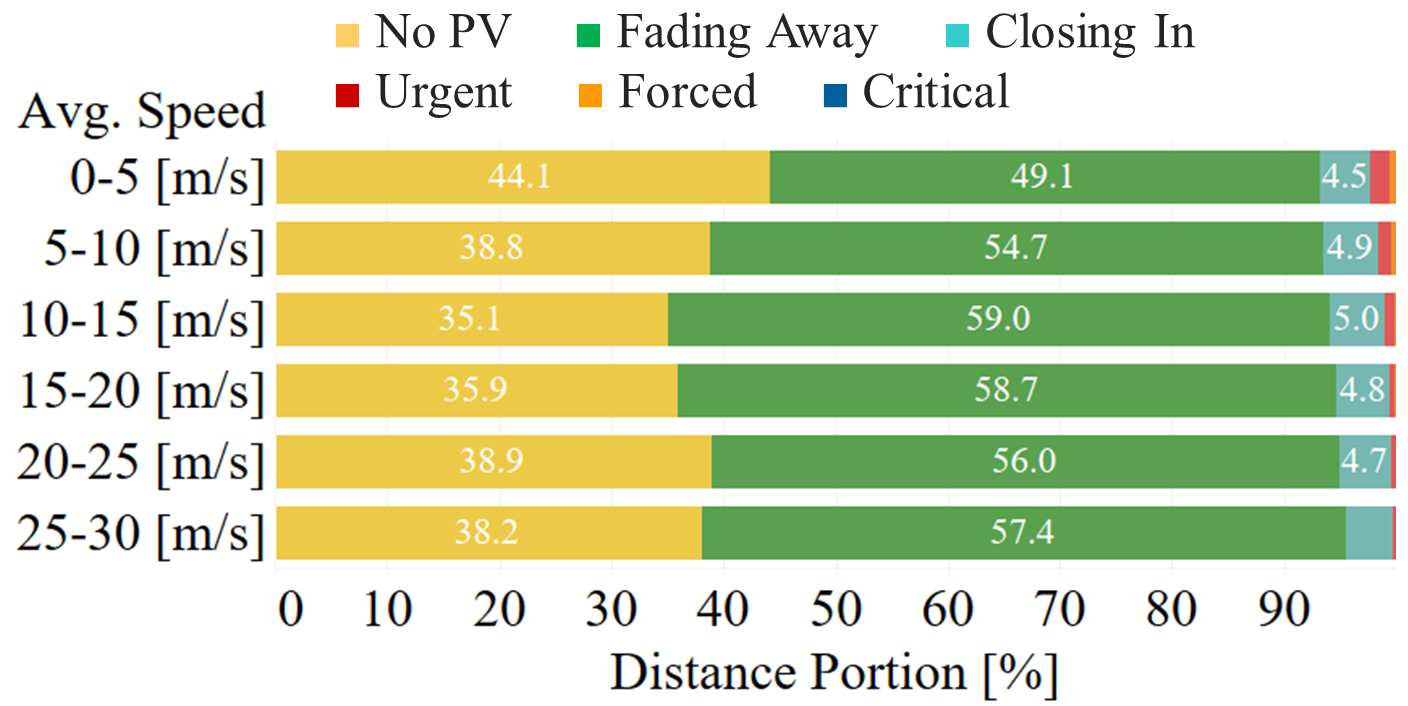} 
    \caption{Collision risk level distribution depending on the average speed range.}
    \label{fig:trip collision risk level}
\end{figure}

In addition to analyzing trip characteristics, we also analyzed the trip composition of: 1) six scenario types and 2) six collision risk levels. The six collision risk levels are defined based on time to collision (TTC), as follows:
\begin{itemize}[noitemsep]
    \item No preceding vehicle (PV): no signals of PV
    \item Fading Away: $v-v_\mathrm{p}\leq0$  
    \item Closing In: $\mathrm{TTC}\geq5.5 \ \mathrm{s}$ and $v-v_\mathrm{p}>0$ 
    \item Urgent: $3 \ \mathrm{s} \leq \mathrm{TTC} < 5.5 \ \mathrm{s}$ and $v-v_\mathrm{p}>0$
    \item Forced:  $1 \ \mathrm{s} \leq \mathrm{TTC} < 3 \ \mathrm{s}$ and $v-v_\mathrm{p}>0$
    \item Critical:  $\mathrm{TTC} < 1 \ \mathrm{s}$ and $v-v_\mathrm{p}>0$
\end{itemize}
Note that the TTC is defined by the time that it takes to collide with the preceding vehicle assuming that relative speed is held, $\mathrm{TTC}=(s_\mathrm{p} - s - l_\mathrm{p})/\max(0, v - v_\mathrm{p})$, where $s$, $v$, and $l$ are position, speed, and vehicle length, respectively, and a subscript $p$ indicates the preceding vehicle. In Figures \ref{fig:trip scenario distribution} and \ref{fig:trip collision risk level}, we averaged the distance portion of a total over all trips corresponding to the average speed range. 

Figure \ref{fig:trip scenario distribution} shows that scenario type distribution differs with trip types.  
Trips with higher average speeds involve longer Crs scenarios, which is the primary reason for longer traveled distance.
In trips with lower average speeds, scenarios involving acceleration and braking caused by road braking events become a major portion of trip, compared to the Crs scenarios.

As shown in Figure \ref{fig:trip collision risk level}, it is evident that the six collision risk levels are not correlated with trip types. Generally, the distance portion of the No PV level is about 35\% to 44\%, while that of the other levels describing car-following scenarios, where a preceding vehicle coexists, is about 56\% to 65\%. 
In car-following scenarios, the Fading Away level is the largest portion of about 49\% to 59\%, followed by the Closing In level at about 4\% to 5\%. The portion of safety-critical levels, including the Urgent, Forced, and Critical levels, is below 3\%.  

\begin{figure}[h!]
    \centering
    \includegraphics[width=0.6\columnwidth,keepaspectratio]{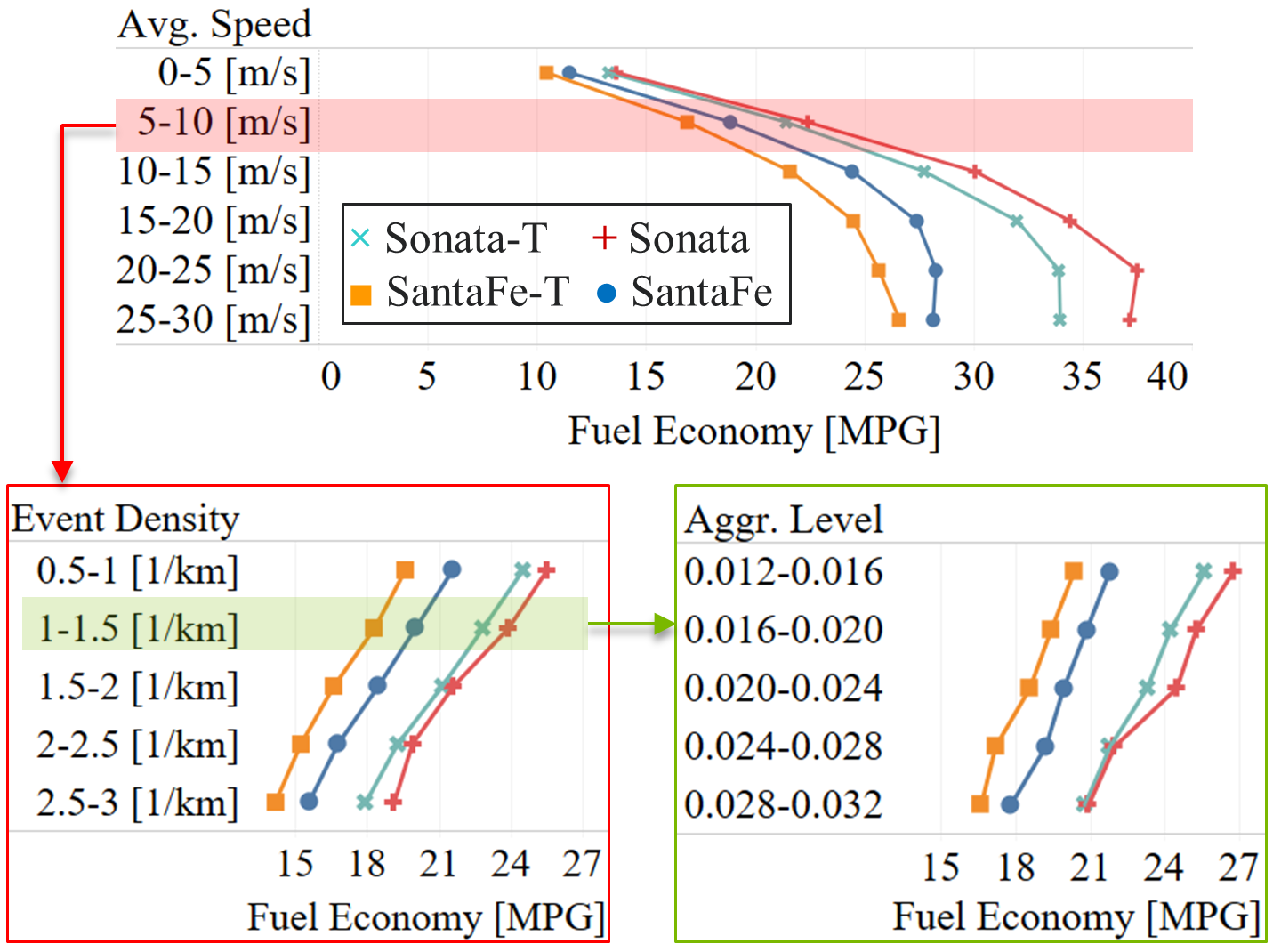} 
    \caption{Average fuel economy analysis: 1) average speed range and vehicle type (top), 2) road event density and vehicle type for the given average speed range of 5--10 m/s (bottom left), and 3) driver aggressiveness level and vehicle type for the given average speed range of 5--10 m/s and event density of 1--1.5 l/km (bottom right).}
    \label{fig:trip fuel economy}
\end{figure}

We computed the fuel economy, the unit of which is mile per gallon (MPG), for each trip using the instantaneous fuel flow rate signal provided in the dataset. 
As shown in Figure \ref{fig:trip fuel economy}, fuel economy varies depending on two external factors (i.e., average speed and braking event density) and two internal factors (i.e., vehicle model and driver aggressiveness level). Note that aggressiveness level is defined as the acceleration energy per kilometer, by integrating the squared acceleration signal over time and dividing it by the traveled distance.
Fuel economy tends to improve with higher average speeds and lower event densities. 
This trend is aligned with the standard fuel economy tests that highlight lower fuel economy in city driving compared to highway driving. 
Within the same average speed range, braking events during the trip play a key role in determining fuel economy. Acceleration is necessary to reach the desired speed after braking, resulting in increased fuel consumption. Furthermore, even with the same braking event density, the driving style of individuals affects fuel economy, with more aggressive driving resulting in lower fuel economy. As expected, larger vehicles equipped with turbocharged engines show lower fuel economy. 
This fuel economy analysis highlights the importance of designing trip scenarios for evaluating fuel economy performance.  

\subsection{Scenario-level insights}
The scenario-level of analysis examines a series of scenarios composed during individual, real-world, daily trips. Conditions within each scenario can be extracted to establish their realistic ranges, which are related to real-world road characteristics. Specifically, for braking scenarios (i.e., BSnA and BnA), we considered the following three scenario parameters:
\begin{itemize}[noitemsep]
    \item Approaching speeds: speeds at the beginning of scenarios.
    \item Perceivable distance to road event: remaining distance to the braking road event at the beginning of scenarios.
    \item Curvature for turning: maximum curvature near the braking event.
\end{itemize}
\begin{figure}[h!]
    \centering
    \includegraphics[width=0.5\columnwidth,keepaspectratio]{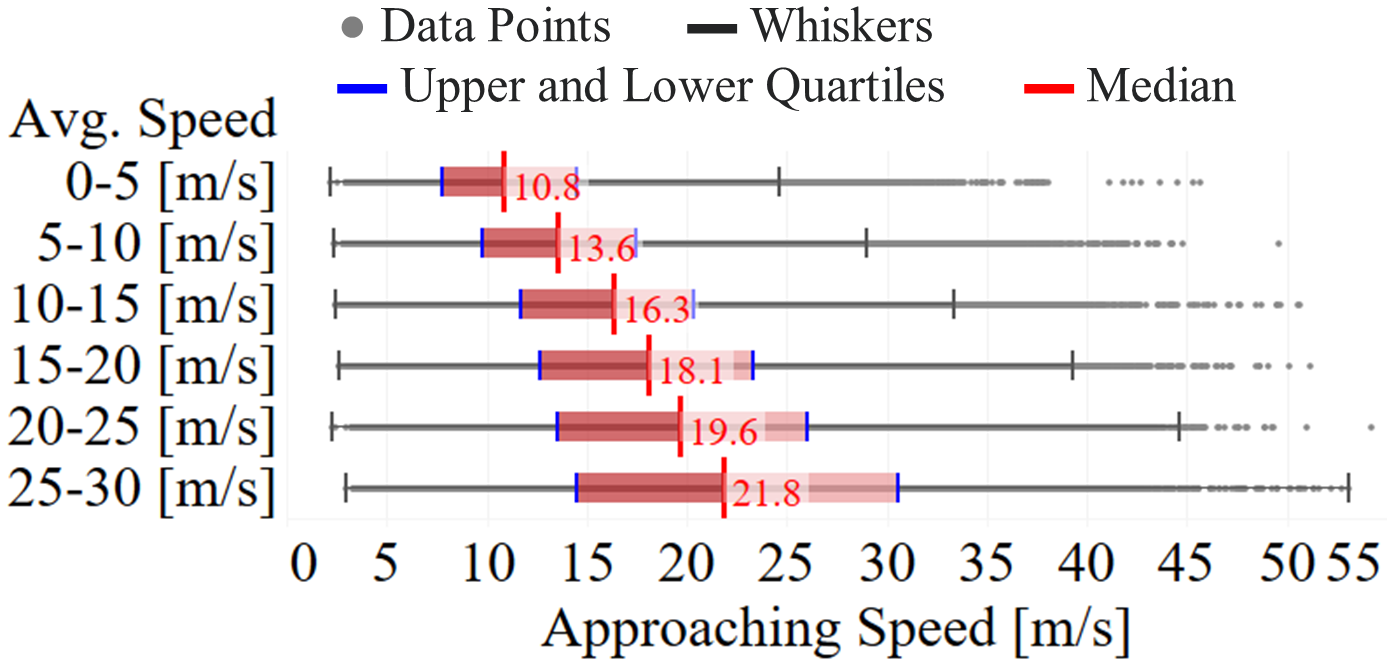} 
    \caption{Approaching speed depending on the average speed range.}
    \label{fig:scenario approaching speed}
\end{figure}
The average speed range is related to the maximum speed limit on roads. Thus, its correlation with the median approaching speed is evident, as shown in Figure \ref{fig:scenario approaching speed}. Braking scenarios with low approaching speeds are present even in the high average speed range, as low-speed driving, as well as stops near origins and destinations, are necessary. Also, higher approaching speeds than the average speed can occur due to various combinations of road types in a single trip (e.g., 80\% city + 20\% highway roads).  

\begin{figure}[h!]
    \centering
    \includegraphics[width=0.5\columnwidth,keepaspectratio]{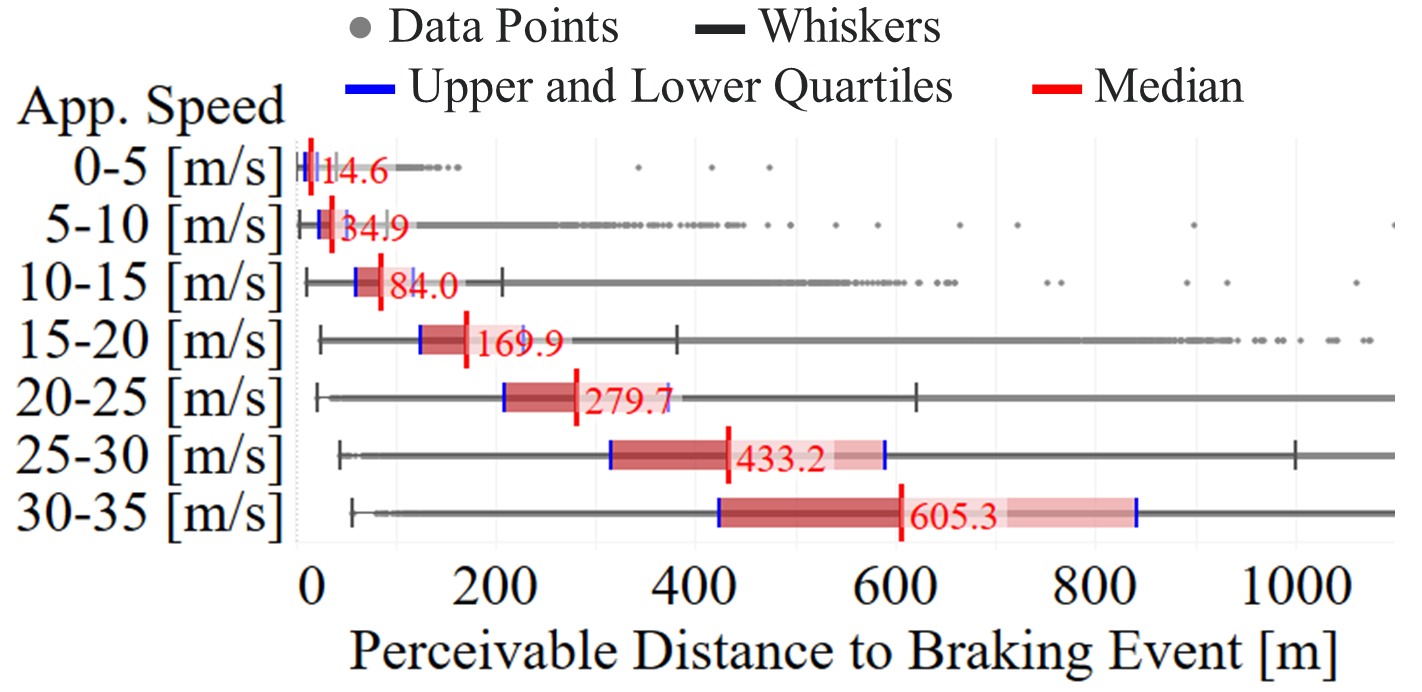} 
    \caption{Perceivable distance to braking event depending on the approaching speed range.}
    \label{fig:scenario distance to event}
\end{figure}
Figure \ref{fig:scenario distance to event} shows that a higher approaching speed range resulted in longer median perceivable distances, given the requirement that roads should be designed to guarantee sufficient braking distance for the maximum speed limit. Short perceivable distances can occur even in the high approaching speed range due to sudden changes in road events (e.g., the upcoming traffic light switching from green to yellow). Individual preferences also influence perceivable distances; for example, some individuals may prefer longer distances compared to medium ones.  

\begin{figure}[h!]
    \centering
    \includegraphics[width=0.5\columnwidth,keepaspectratio]{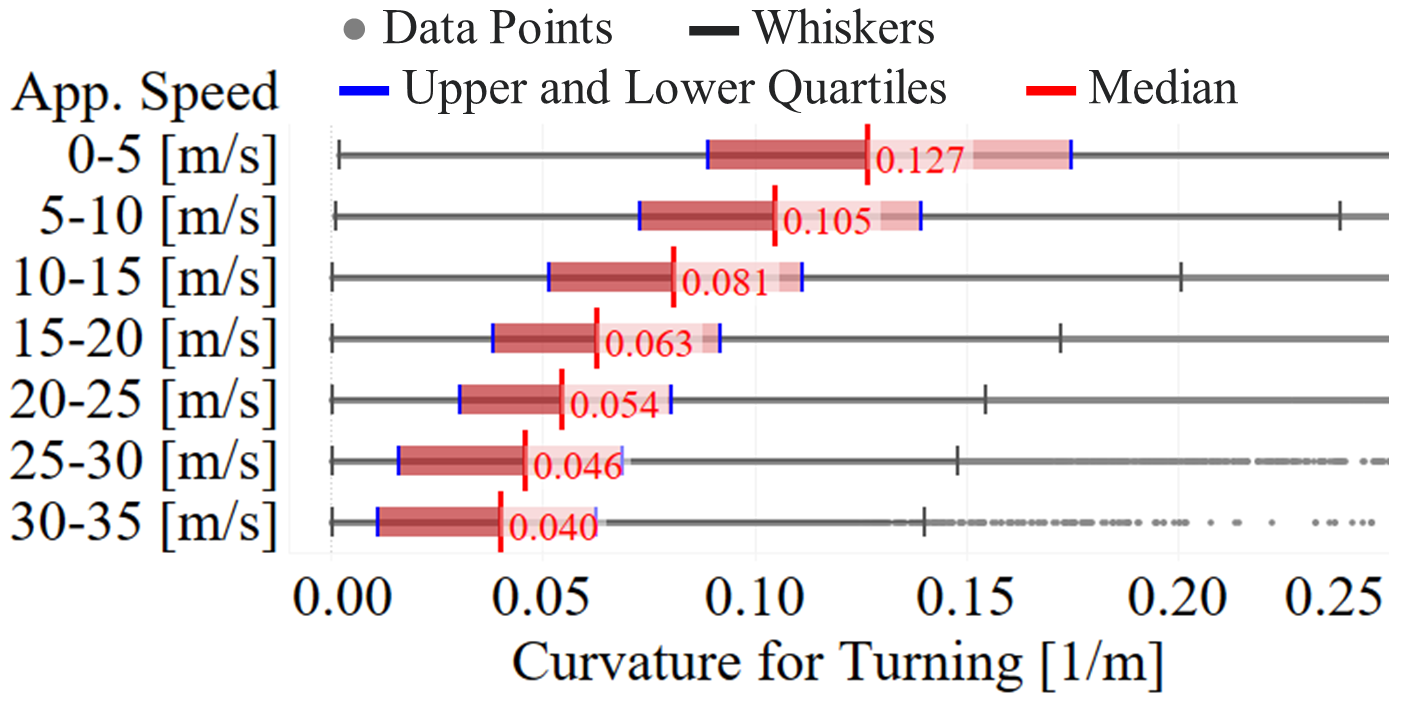} 
    \caption{Curvature for turning depending on the approaching speed range.}
    \label{fig:scenario curvature}
\end{figure}
Assuming steady-state cornering, we computed curvature ($k$) signal using speed ($v$) and yaw rate ($r$) signals, defined by $k=r/v$. We considered only turning-involved braking scenarios, which are identified by the absolute value of the yaw rate signal being above 5 deg/s, to examine the upcoming road curvature for the given approaching speed range. Figure \ref{fig:scenario curvature} shows that the median road curvatures generally decrease with approaching speeds, and the trend in the upper whisker aligns for safe turning purposes in road design.

\begin{figure}[h!]
    \centering
    \includegraphics[width=0.6\columnwidth,keepaspectratio]{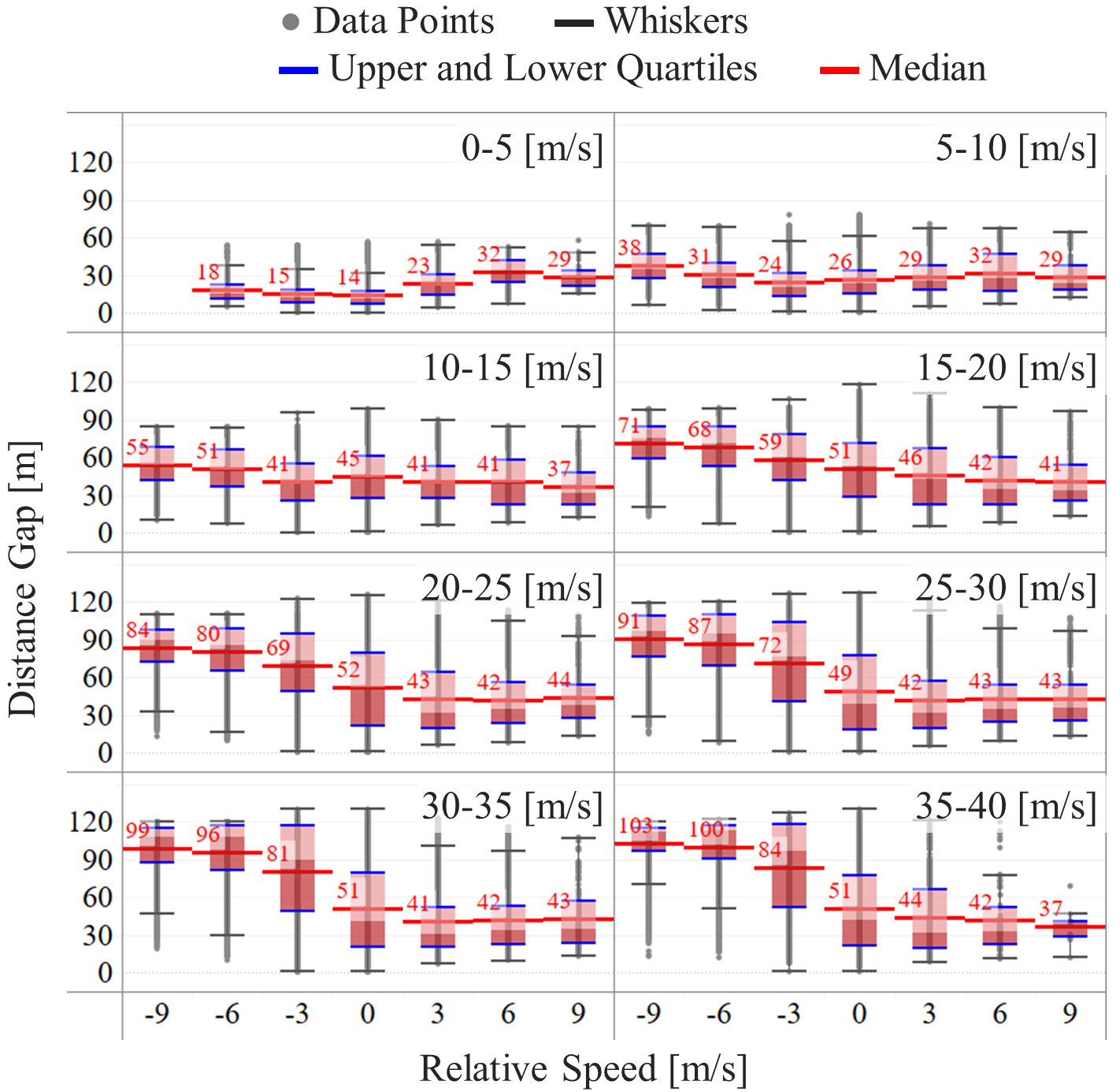} 
    \caption{Distance gap (on the y-axis) depending on relative speed (on the x-axis) and the approaching speed range (in each cell).}
    \label{fig:Scenario gap and relative speed}
\end{figure}
For safety test purposes, analyzing the conditions of PV appearances in cut-in/lane-change scenarios is important. We considered three scenario parameters: 1) distance gap ($s_\mathrm{p}-s$), 2) relative speed ($v_\mathrm{p}-v$), and 3) approaching speed ($v$). Figure \ref{fig:Scenario gap and relative speed} shows the distance gap, relative speed, and speed at the moment the PV was detected right after its absence. Negative and positive relative speeds indicate scenarios of closing in and fading away, respectively. For closing-in scenarios, the median distance gap increases with higher approaching speed, and it also increases with lower relative speed for each approaching speed range. On the other hand, for fading-away scenarios, the median distance gap shows no distinct trend. The lower whiskers of the distance gap at the lowest relative speed (-9 m/s) can be used for the most risky closing-in test scenarios, depending on the approaching speed range, which automated vehicles must guarantee for collision-free driving with human-acceptable comforts.

\subsection{Driving-level insights}
We considered three driving regimes, such as acceleration, braking, and coasting regimes, for the driving level of analysis. These regimes are obtained through further trajectory segmentation in the driving-level process for BSnA, BnA, A, and B scenarios. 

\begin{figure}[h!]
    \centering
    \includegraphics[width=0.5\columnwidth,keepaspectratio]{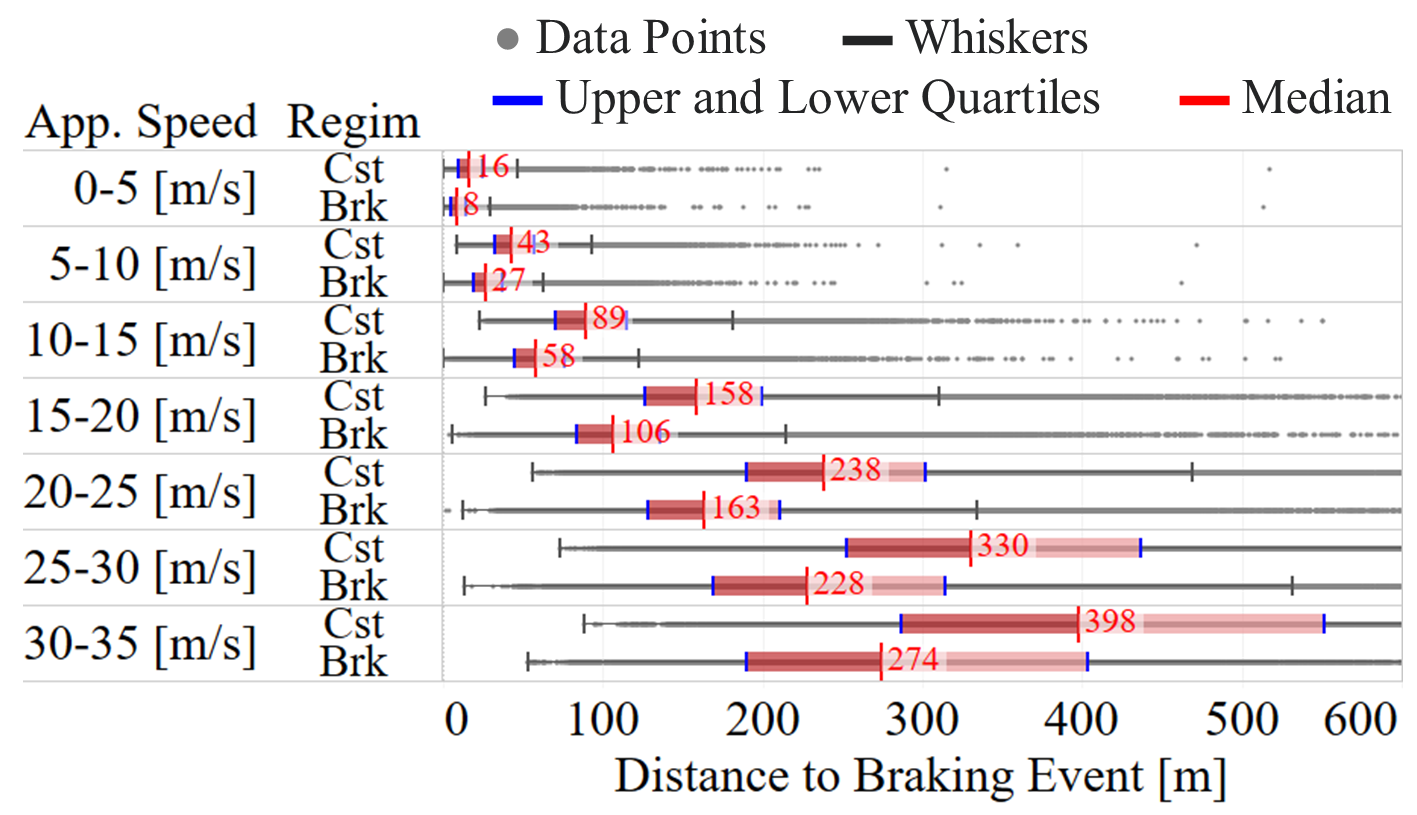} 
    \caption{Distance to the braking event depending on the approaching speed range at the initiation of both coasting and braking regimes.}
    \label{fig:driving decision making}
\end{figure}

We first analyzed decision-making characteristics regarding when to initiate coasting and braking regimes. Figure \ref{fig:driving decision making} shows the distance to the braking event, defined by the remaining distance to the braking road event at which coasting and braking regimes initiate, respectively. 
To slow down for upcoming braking events, drivers initiate the coasting regime first, followed by the braking regime. The initiation rates for the coasting regime are about 42.9\% (68,594 out of 159,999), 39.9\% (166,164 out of 416,133), and 51.3\% (135,509 out of 264,103) for B, BnA, and BSnA scenarios, respectively. 
Higher approaching speeds result in earlier reactions to upcoming braking events within perceivable distances, leading drivers to initiate coasting and braking regimes earlier. The range between the lower and upper quartiles becomes wider.
Two distance curves can serve as thresholds for regime decision-making. For example, at the approaching speed range of 30--35 m/s, a driver initiates coasting and braking regimes at about 398 m and 274 m before the braking event, respectively.

Then, we analyzed the characteristics of real-world human driving behaviors by examining median values that represent typical patterns from the following perspectives:
\begin{itemize}[noitemsep]
    \item Distance and time: the distance traveled and the time taken to reach the final speed ($v_\mathrm{f}$) from the initial speed ($v_\mathrm{0}$).
    \item Aggressiveness level (as defined in Section \ref{subsec:trip}).
    \item Turning speed: the minimum speed near the braking road event when turning.
\end{itemize}

\begin{figure}[h!]
    \centering
    \includegraphics[width=0.55\columnwidth,keepaspectratio]{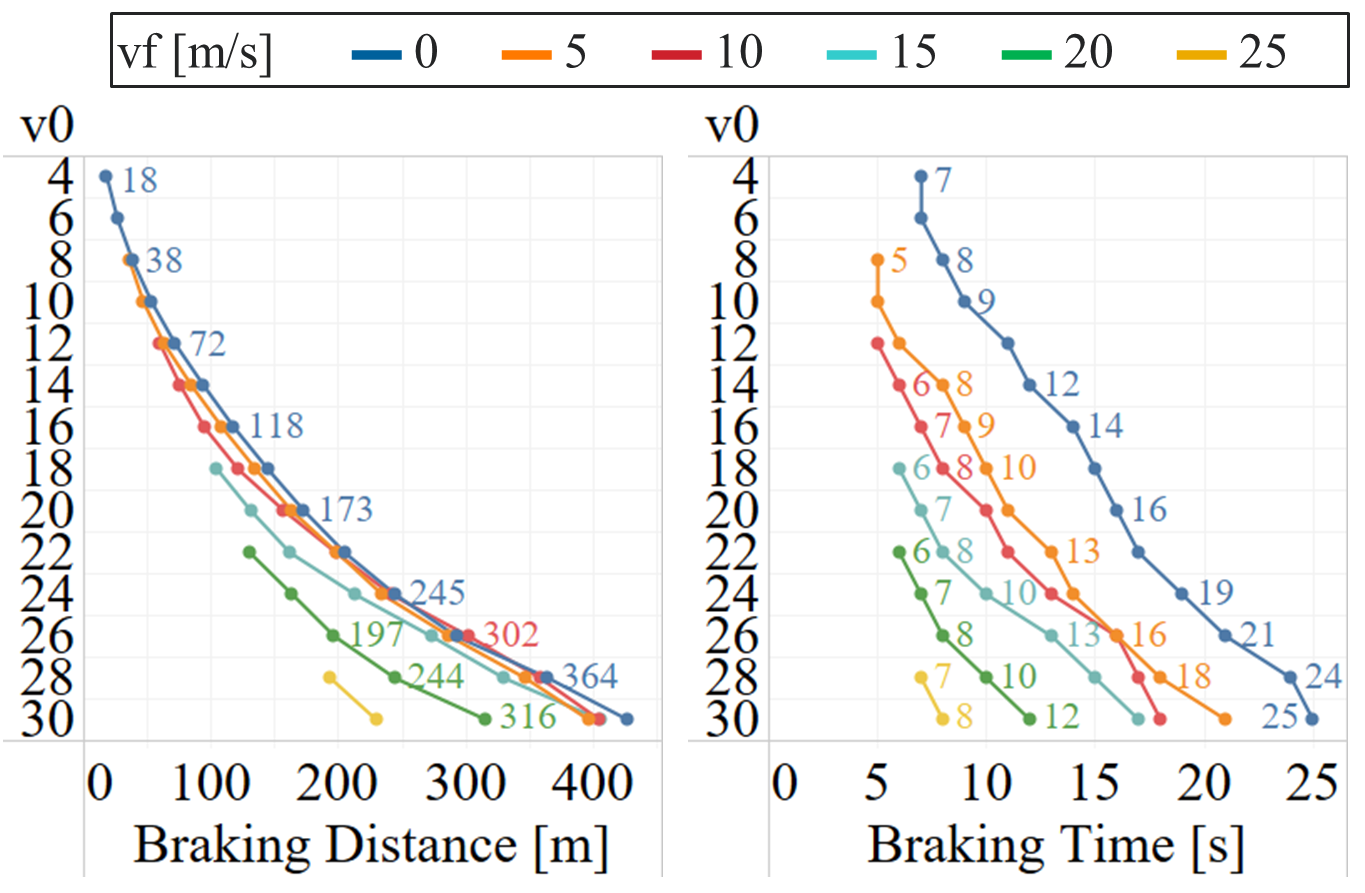} 
    \caption{Median braking distance (left) and time (right) depending on initial speeds for various final speeds.}
    \label{fig:driving braking distance and time}
\end{figure}

Figure \ref{fig:driving braking distance and time} shows median distance and time during the braking regimes in ``B''-involved scenarios. In braking regimes, the final speed is the speed that results from the conclusion of the braking event (e.g., a green light is turned on after a red light). 
Higher initial speeds resulted in longer braking distances and times to reach a given final speed. 
In general, increased braking time led to an increased braking distance. 
When braking from the same initial speed, the extent of the increase in braking time is nearly constant with the decrease in final speed. Thus, a lower final speed (i.e., a higher speed decrease) increased both braking time and distance. 

\begin{figure}[h!]
    \centering
    \includegraphics[width=0.55\columnwidth,keepaspectratio]{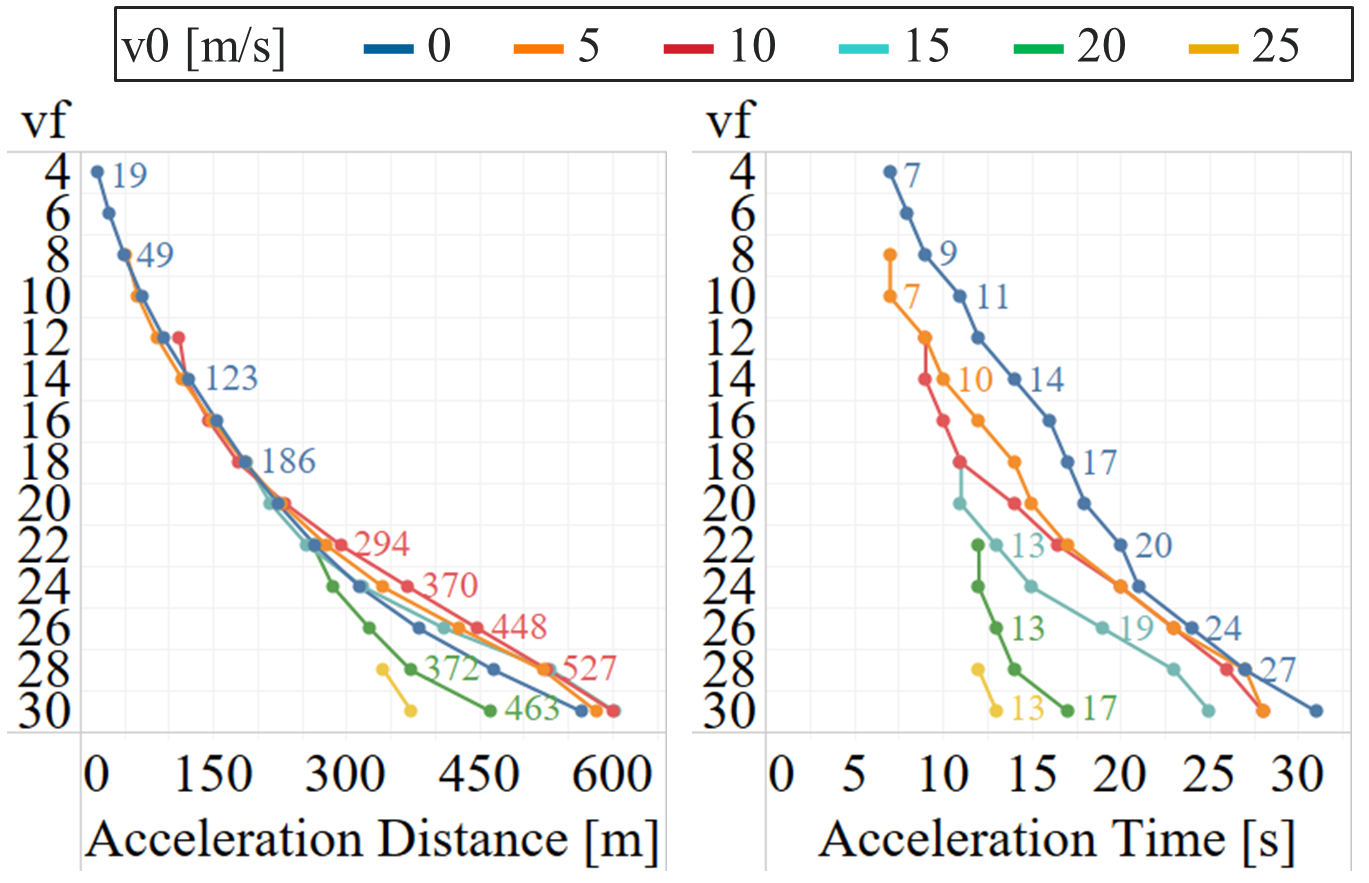} 
    \caption{Median acceleration distance (left) and time (right) depending on final speeds for various initial speeds.}
    \label{fig:driving acceleration distance and time}
\end{figure}

Figure \ref{fig:driving acceleration distance and time} shows median distance and time during the acceleration regimes in ``A''-involved scenarios. 
In acceleration regimes, the final speed should be set as desired by the drivers, considering the maximum speed limits on roads. 
Similar to Figure \ref{fig:driving braking distance and time}, higher final speeds resulted in longer acceleration distances and times.   
It is worth noting that drivers tended to have stronger acceleration levels when accelerating from a low-speed range (e.g., $v_0<10$ m/s) to a high-speed range (e.g., $v_\mathrm{f}>25$ m/s). These stronger acceleration levels led to shorter acceleration times; thus, the acceleration distance for low initial speeds may be shorter than that for medium initial speeds (e.g., $v_0= 0 \ \mathrm{vs.} \ 10$ m/s). 

\begin{figure}[h!]
    \centering
    \includegraphics[width=0.55\columnwidth,keepaspectratio]{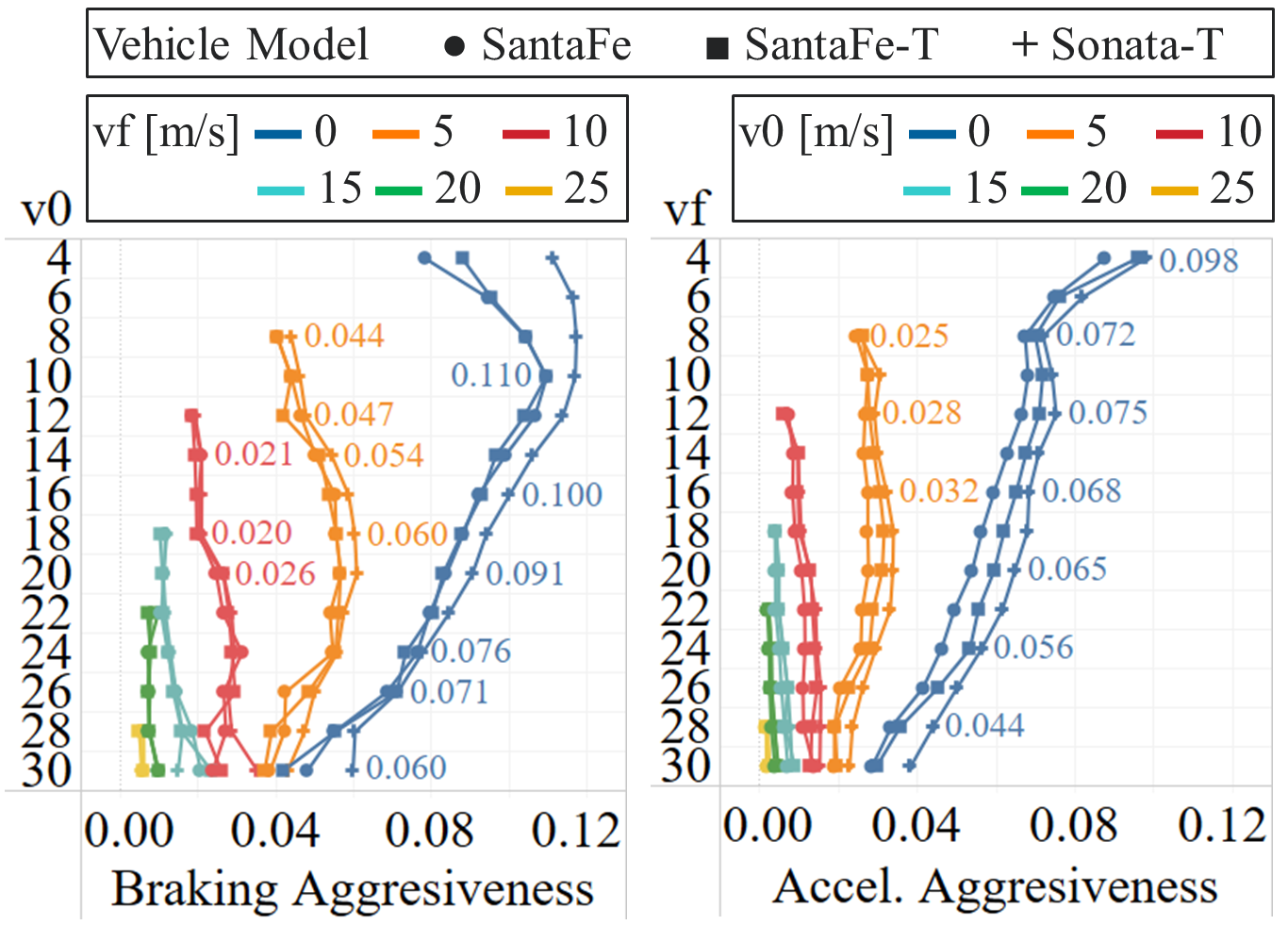} 
    \caption{Median aggressiveness levels for braking (left) and acceleration (right) depending on initial and final speeds for three vehicles models.}
    \label{fig:driving aggressiveness levels}
\end{figure}

As shown in Figure \ref{fig:driving aggressiveness levels}, the aggressiveness level decreases with an increase in the lower speed, i.e., final and initial speeds for braking and acceleration regimes, respectively, due to the decrease in speed decrease/increase. In this figure, we excluded Sonata data due to its relatively low volume size, as shown in Table \ref{tab:data summary}. It is worth noting that  in scenarios involving braking to stops and accelerating from stops, braking from lower speeds decreases the aggressiveness level, while accelerating from lower speeds tends to increase it. Generally speaking, it is observed that human drivers prefer more comfortable stops and more aggressive acceleration within a range of low speeds. 
The aggressiveness level was also influenced by vehicle models, indicating that the smallest vehicle size with the turbo engine (Sonata-T) resulted in the most aggressive behaviors. The vehicle size appears to be the only factor affecting braking aggressiveness level. 

\begin{figure}[h!]
    \centering
    \includegraphics[width=0.55\columnwidth,keepaspectratio]{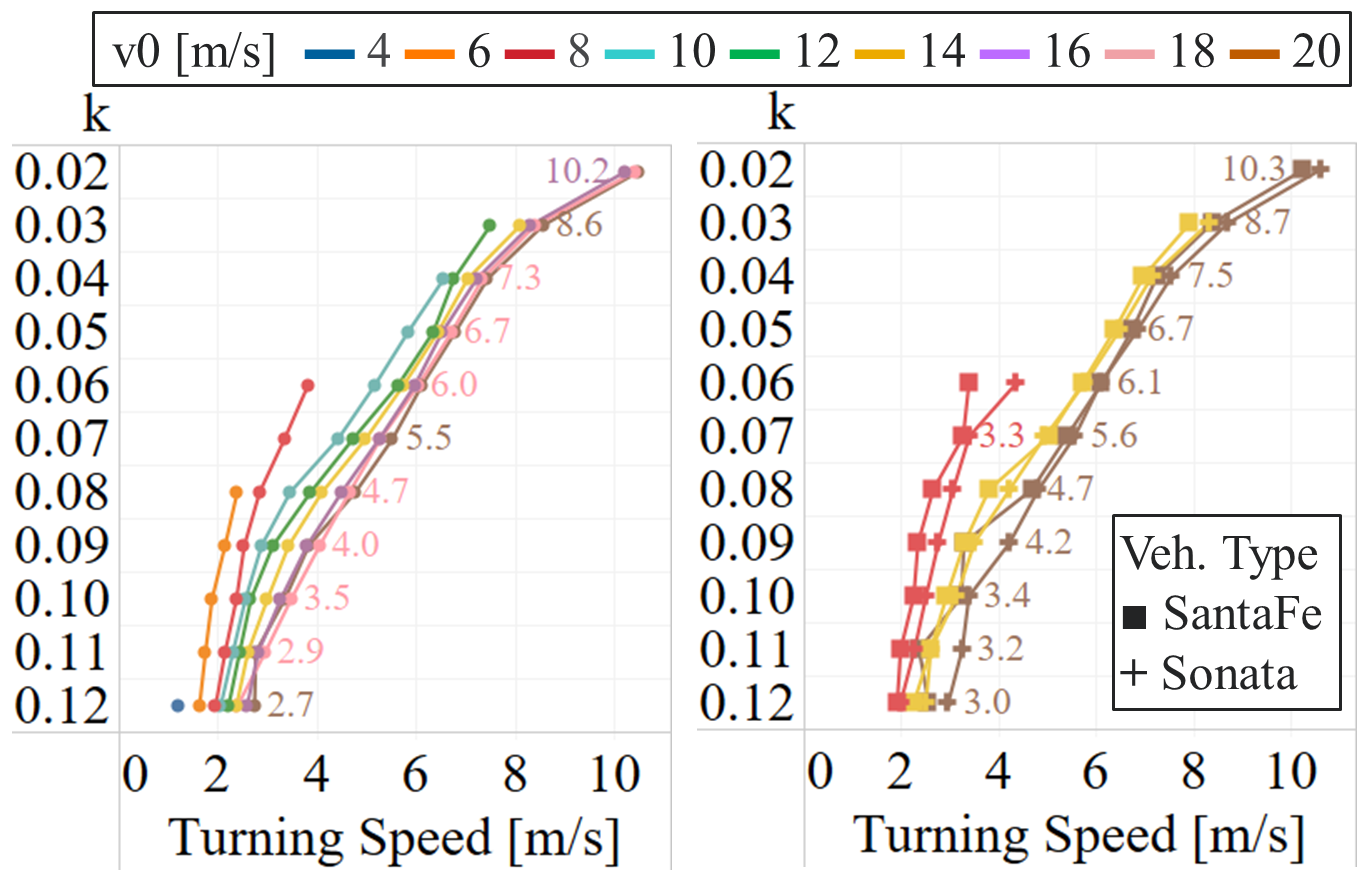} 
    \caption{Median turning speeds depending on curvature for various initial speeds (left) and selected turning speed curves for two vehicle types (right).}
    \label{fig:driving turning speeds}
\end{figure}

A driver’s intention for a turn can be identified by checking whether the initial speeds exceed the maximum turning speeds.
Note that the maximum turning speed curve is constructed by fitting a curve to selected speed data points, where 99.9\% of the data lies below these points at specific curvatures.
As shown in Figure \ref{fig:driving turning speeds}, drivers reduce the speed more to make turns at higher curvatures. These turning speeds also depend on initial speeds, indicating that lower initial speeds result in lower turning speeds even for the same curvature. 
Three initial speeds (8, 14, and 20 m/s) were selected to show the impact of vehicle type on turning speeds. Drivers in bigger vehicles exhibit more caution while turning, reducing speeds more at the same curvature. 

\section{Conclusion}\label{sec: conclusion}

In this paper, we presented a comprehensive data analytics framework for processing and exploring a large Hyundai customer driving dataset. This framework allows us to extract insights at three levels: trip, scenario, and driving behavior. 
Trip-level insights support the design of real-world trip tests to evaluate fuel economy by ensuring statistical trip properties. Deploying various realistic trip tests enables more accurate and reliable evaluations of the fuel economy of CAVs. 
Insights at the scenario level can be readily applied in designing individual test scenarios to validate functionality. Descriptive statistics provide various realistic values for setting scenario parameters, enabling the evaluation of braking-to-stop performance in braking scenarios, collision-avoidance performance in cut-in/lane-change scenarios, and both performances in combined scenarios. Scenarios can range from typical to safety-critical based on the initial parameter setup. 
Insights at the behavior level can be used to model human drivers and establish guidelines for human-like automated driving. We highlighted general trends in realistic human driving behaviors, focusing on typical patterns. This includes answering the following questions: at what distance does a driver initiate coasting and braking? What factors impact braking and acceleration behaviors? At what speed does a driver drive for turning? 

In the future, we aim to develop and validate a human driver model using this dataset. Additionally, we plan to deploy the model in simulations and generate baseline results for test scenarios designed by data-driven parameters for a fair comparison with CAV results.

\section*{Disclaimer}
The submitted manuscript has been created by UChicago Argonne, LLC, Operator of Argonne National Laboratory (``Argonne''). Argonne, a U.S. Department of Energy Office of Science laboratory, is operated under Contract No. DEAC02-06CH11357. The U.S. Government retains for itself, and others acting on its behalf, a paid-up nonexclusive, irrevocable worldwide license in said article to reproduce, prepare derivative works, distribute copies to the public, and perform publicly and display publicly, by or on behalf of the Government. The Department of Energy (DOE) will provide public access to these results of federally sponsored research in accordance with the DOE Public Access Plan.








\appendix
\section{Pseudo codes} \label{App: codes}
\begin{algorithm}[h!]
    \KwInput{A time series of driving data in discrete time $t_k$ for $k=0 \cdots N$ ($v_k$ is speed at time $k$, where the subscript $k$ denotes discrete time).}
    \KwOutput{Times of local extremes ($\mathbf{k}_{\text{min}}$ and $\mathbf{k}_{\text{max}}$) and influential local extremes ($\mathbf{k}_{\text{inf,min}}$ and $\mathbf{k}_{\text{inf,max}}$).}
    find local minimums, $\mathbf{k}_{\text{min}}\in [0, N]$, and local maximums, $\mathbf{k}_{\text{max}}\in [0, N]$.\\
    initialization, $\mathbf{k}_{\text{inf,min}} \leftarrow \pmb{\emptyset} $, $\mathbf{k}_{\text{inf,max}} \leftarrow \pmb{\emptyset} $, and $k_{\text{inf,min}}^{\mathrm{prv}} \leftarrow 0$ \\
    \For {each, $k_{\mathrm{min},i} \in \mathbf{k}_{\mathrm{min}}$} {
        $\mathbf{k}_{\text{inf,max}}^{\mathrm{slc}} \leftarrow \{k_{\mathrm{max}}\in \mathbf{k}_{\text{max}} |k_{\mathrm{max}}\in [k_{\text{inf,min}}^{\mathrm{prv}}, N], v_{k_{\text{max}}} - v_{k_{\text{min}}^{\mathrm{nxt}}} > 5 \mathrm{m/s, where} \ k_{\text{min}}^{\mathrm{nxt}}=\arg \min_{k_{\text{min}} > k_{\text{max}}} {k_{\text{min}}} \}$ \\
        \If {$\exists k_{\mathrm{max}} \in \mathbf{k}_{\mathrm{inf,max}}^{\mathrm{slc}}: k_{\mathrm{max}} < k_{\mathrm{min},i} $ } {
            $k_{\mathrm{max,f}} \leftarrow \arg \min_{k_{\mathrm{max}}\in \mathbf{k}_{\mathrm{inf,max}}^{\mathrm{slc}}, \ k_{\mathrm{max}} > k_{\mathrm{min},i}} {k_{\mathrm{max}}} $ \\
            $\mathbf{k}_{\text{inf,min}}^{\mathrm{slc}} \leftarrow \{k_{\mathrm{min}}\in \mathbf{k}_{\text{min}} | k_{\mathrm{min}}\in [k_{\text{min},i}, k_{\mathrm{max,f}}], v_{k_{\text{max}}^{\mathrm{nxt}}} - v_{k_{\text{min}}} > 5 \mathrm{m/s, where} \ k_{\text{max}}^{\mathrm{nxt}}=\arg \min_{k_{\text{max}} > k_{\text{min}}} {k_{\text{max}}} \}$\\
            \If {$\mathbf{k}_{\mathrm{inf,min}}^{\mathrm{slc}}$ is not the empty set} {
                $v_{\mathrm{min}}\leftarrow \min(v_{\mathbf{k}_{\mathrm{inf,min}}^{\mathrm{slc}}})$\\
                \If {$v_{{k}_{\mathrm{min},i}}\leq v_{\mathrm{min}}$ and $v_{\mathrm{min}} \neq 0$} {
                    $k_{\mathrm{inf,min}}^{\mathrm{slc}}\leftarrow k_{\mathrm{min},i}$ \\
                    $k_{\mathrm{inf,max}}^{\mathrm{slc}}\leftarrow \arg \max_{k_{\mathrm{max}} \in \mathbf{k}_{\mathrm{max}}^{\mathrm{slc}}} {v_{k_{\mathrm{max}}} }$, where $\mathbf{k}_{\mathrm{max}}^{\mathrm{slc}}=\{k_{\mathrm{max}}\in \mathbf{k}_{\mathrm{max}} | k_{\mathrm{max}}\in [k_{\text{inf,min}}^{\mathrm{prv}}, k_{\mathrm{inf,min}}^{\mathrm{slc}}] \}$ \\                    
                    $\mathbf{k}_{\text{inf,min}} \leftarrow \mathbf{k}_{\text{inf,min}} \cup \{ k_{\mathrm{inf,min}}^{\mathrm{slc}} \}$, $\mathbf{k}_{\text{inf,max}} \leftarrow \mathbf{k}_{\text{inf,max}} \cup \{ k_{\mathrm{inf,max}}^{\mathrm{slc}} \}$, and $k_{\text{inf,min}}^{\mathrm{prv}} \leftarrow k_{\mathrm{inf,min}}^{\mathrm{slc}}$
                }
            }
        } 
    }      
    return $\mathbf{k}_{\text{min}}$, $\mathbf{k}_{\text{max}}$, $\mathbf{k}_{\text{inf,min}}$ and $\mathbf{k}_{\text{inf,max}}$  \\      
\caption{Road event timing search}
\label{alg1}
\end{algorithm}
\begin{algorithm}[h!]
    \KwInput{$\mathbf{k}_{\text{min}}$, $\mathbf{k}_{\text{max}}$, $\mathbf{k}_{\text{inf,min}}$ and $\mathbf{k}_{\text{inf,max}}$.}
    \KwOutput{Time intervals of each scenario ($\mathbf{k}_{\text{0}}$ and $\mathbf{k}_{\text{f}}$) and scenario type ($\mathbf{s}$).}
    initialization, $\mathbf{k}_{\text{0}} \leftarrow \pmb{\emptyset} $,  $\mathbf{k}_{\text{f}} \leftarrow \pmb{\emptyset}$, and $\mathbf{s} \leftarrow \pmb{\emptyset} $ \\    
    \For {each, $k_{\mathrm{inf,min},i} \in \mathbf{k}_{\mathrm{inf,min}}$, and each, $k_{\mathrm{inf,max},i} \in \mathbf{k}_{\mathrm{inf,max}}$} {
        $\mathbf{k}_{\text{inf,max}}^{\mathrm{brk}} \leftarrow \{k_{\mathrm{max}}\in \mathbf{k}_{\text{max}} |k_{\mathrm{max}}\in [k_{\mathrm{inf,max},i}, k_{\mathrm{inf,min},i}], v_{k_{\text{max}}} - v_{k_{\text{min}}^{\mathrm{nxt}}} > 5 \mathrm{m/s, where} \ k_{\text{min}}^{\mathrm{nxt}}=\arg \min_{k_{\mathrm{min}}\in \mathbf{k}_{\text{min}}, k_{\text{min}} > k_{\text{max}}} {k_{\text{min}}} \}$ \\
        $k_\mathrm{f}^{\mathrm{slc}}\leftarrow \emptyset$ \\         
        \For {each,         ${k}_{\mathrm{inf,max},i}^{\mathrm{brk}}\in\mathbf{k}_{\mathrm{inf,max}}^{\mathrm{brk}}$}{
            ${k}_{\mathrm{min}}^{\mathrm{nxt}} \leftarrow \arg \min_{k_{\mathrm{min}}\in \mathbf{k}_{\text{min}},  {k}_{\mathrm{min}}>{k}_{\mathrm{inf,max},i}^{\mathrm{brk}}} {{k}_{\mathrm{min}}}$ \\
            \If{$t_{{k}_{\mathrm{inf,max},i+1}^{\mathrm{brk}}} - t_{{k}_{\mathrm{min}}^{\mathrm{nxt}}} > 6\mathrm{s}$} {
                $k_\mathrm{0}^{\mathrm{slc}}\leftarrow {k}_{\mathrm{inf,max},i}^{\mathrm{brk}}$, $k_\mathrm{f}^{\mathrm{slc}}\leftarrow {k}_{\mathrm{min}}^{\mathrm{nxt}}$, and $s^{\mathrm{slc}}\leftarrow\{ \text{``B''} \}$ \\
                $\mathbf{k}_{\mathrm{0}} \leftarrow \mathbf{k}_{\mathrm{0}} \cup \{ k_{\mathrm{0}}^{\mathrm{slc}} \}$, $\mathbf{k}_{\mathrm{f}} \leftarrow \mathbf{k}_{\mathrm{f}} \cup \{ k_{\mathrm{f}}^{\mathrm{slc}} \}$, and $\mathbf{s} \leftarrow \mathbf{s} \cup s^{\mathrm{slc}}$
            }            
        }
        \eIf{$k_\mathrm{f}^{\mathrm{slc}}$ is not the empty set} {
            $k_{\mathrm{0}}^{\mathrm{slc}}\leftarrow \arg \min_{ {k}_{\mathrm{max}}\in \mathbf{k}_{\text{inf,max}}^{\mathrm{brk}}, {k}_{\mathrm{max}}>k_\mathrm{f}^{\mathrm{slc}}} {{k}_{\mathrm{max}}}$ \\        
        }
        {
            $k_{\mathrm{0}}^{\mathrm{slc}}\leftarrow \mathbf{k}_{\text{inf,max}}^{\mathrm{brk}}\{1\}$ \\        
        }
        \eIf{$v_{k_{\mathrm{inf,min},i}}=0$} {
            $s^{\mathrm{mdl}}\leftarrow \{ \text{``BSnA''} \}$  
        }
        {
            $s^{\mathrm{mdl}}\leftarrow \{ \text{``BnA''} \}$  
        }  
        $\mathbf{k}_{\text{inf,min}}^{\mathrm{acc}} \leftarrow \{k_{\mathrm{min}}\in \mathbf{k}_{\text{min}} |k_{\mathrm{min}}\in [k_{\mathrm{inf,min},i}, k_{\mathrm{inf,max},i+1}], v_{k_{\text{max}}^{\mathrm{nxt}}} - v_{k_{\text{min}}} > 5 \mathrm{m/s, where} \ k_{\text{max}}^{\mathrm{nxt}}=\arg \min_{k_{\mathrm{max}}\in \mathbf{k}_{\text{max}}, k_{\text{max}} > k_{\text{min}}} {k_{\text{max}}} \}$ \\
        $k_\mathrm{f}^{\mathrm{slc}}\leftarrow \emptyset$ \\
        \For {each,         ${k}_{\mathrm{inf,min},i}^{\mathrm{acc}}\in\mathbf{k}_{\mathrm{inf,min}}^{\mathrm{acc}}$}{
            ${k}_{\mathrm{max}}^{\mathrm{nxt}} \leftarrow \arg \min_{k_{\mathrm{max}}\in \mathbf{k}_{\text{max}},  {k}_{\mathrm{max}}>{k}_{\mathrm{inf,min},i}^{\mathrm{acc}}} {{k}_{\mathrm{max}}}$ \\
            \If{$t_{{k}_{\mathrm{inf,min},i+1}^{\mathrm{acc}}} - t_{{k}_{\mathrm{max}}^{\mathrm{nxt}}}  > 6\mathrm{s}$} {
                $k_\mathrm{f}^{\mathrm{slc}}\leftarrow {k}_{\mathrm{max}}^{\mathrm{nxt}}$ \\
                \eIf{$i\neq1$}{                 
                    $k_\mathrm{0}^{\mathrm{slc}}\leftarrow {k}_{\mathrm{inf,min},i}^{\mathrm{acc}}$ and $s^{\mathrm{slc}}\leftarrow \{ \text{``A''} \}$              
                }
                {
                    $s^{\mathrm{slc}}\leftarrow s^{\mathrm{mdl}}$                     
                }
                $\mathbf{k}_{\mathrm{0}} \leftarrow \mathbf{k}_{\mathrm{0}} \cup \{ k_{\mathrm{0}}^{\mathrm{slc}} \}$, $\mathbf{k}_{\mathrm{f}} \leftarrow \mathbf{k}_{\mathrm{f}} \cup \{ k_{\mathrm{f}}^{\mathrm{slc}} \}$, and $\mathbf{s} \leftarrow \mathbf{s} \cup s^{\mathrm{slc}}$
            }            
        }
        \If{$k_\mathrm{f}^{\mathrm{slc}}$ is the empty set} {
            $k_{\mathrm{f}}^{\mathrm{slc}}\leftarrow \arg \min_{k_{\mathrm{max}}\in \mathbf{k}_{\text{max}}, {k}_{\mathrm{max}}>\mathbf{k}_{\text{inf,min}}^{\mathrm{acc}}\{\mathrm{n}(\mathbf{k}_{\text{inf,min}}^{\mathrm{acc}})\}} {{k}_{\mathrm{max}}}$ and $s^{\mathrm{slc}}\leftarrow s^{\mathrm{mdl}}$  \\        
             $\mathbf{k}_{\mathrm{0}} \leftarrow \mathbf{k}_{\mathrm{0}} \cup \{ k_{\mathrm{0}}^{\mathrm{slc}} \}$, $\mathbf{k}_{\mathrm{f}} \leftarrow \mathbf{k}_{\mathrm{f}} \cup \{ k_{\mathrm{f}}^{\mathrm{slc}} \}$, and $\mathbf{s} \leftarrow \mathbf{s} \cup s^{\mathrm{slc}}$ \\
        }               
    } 
    return $\mathbf{k}_{\text{0}}$, $\mathbf{k}_{\text{f}}$, and $\mathbf{s}$      
\caption{Slice and dice}
\label{alg2}
\end{algorithm}

\printcredits

\bibliographystyle{cas-model2-names}

\bibliography{Ref}

\begin{thebibliography}{30}
\expandafter\ifx\csname natexlab\endcsname\relax\def\natexlab#1{#1}\fi
\providecommand{\url}[1]{\texttt{#1}}
\providecommand{\href}[2]{#2}
\providecommand{\path}[1]{#1}
\providecommand{\DOIprefix}{doi:}
\providecommand{\ArXivprefix}{arXiv:}
\providecommand{\URLprefix}{URL: }
\providecommand{\Pubmedprefix}{pmid:}
\providecommand{\doi}[1]{\href{http://dx.doi.org/#1}{\path{#1}}}
\providecommand{\Pubmed}[1]{\href{pmid:#1}{\path{#1}}}
\providecommand{\bibinfo}[2]{#2}
\ifx\xfnm\relax \def\xfnm[#1]{\unskip,\space#1}\fi
\bibitem[{Ard et~al.(2023)Ard, Guo, Han, Jia, Vahidi and Karbowski}]{ard_energy-efficient_2023}
\bibinfo{author}{Ard, T.}, \bibinfo{author}{Guo, L.}, \bibinfo{author}{Han, J.}, \bibinfo{author}{Jia, Y.}, \bibinfo{author}{Vahidi, A.}, \bibinfo{author}{Karbowski, D.}, \bibinfo{year}{2023}.
\newblock \bibinfo{title}{Energy-{Efficient} {Driving} in {Connected} {Corridors} via {Minimum} {Principle} {Control}: {Vehicle}-in-the-{Loop} {Experimental} {Verification} in {Mixed} {Fleets}}.
\newblock \bibinfo{journal}{IEEE Transactions on Intelligent Vehicles} \bibinfo{volume}{8}, \bibinfo{pages}{1279--1291}.
\newblock \DOIprefix\doi{10.1109/TIV.2023.3234261}.
\bibitem[{Boschert and Rosen(2016)}]{hehenberger_digital_2016}
\bibinfo{author}{Boschert, S.}, \bibinfo{author}{Rosen, R.}, \bibinfo{year}{2016}.
\newblock \bibinfo{title}{Digital {Twin}—{The} {Simulation} {Aspect}}, in: \bibinfo{editor}{Hehenberger, P.}, \bibinfo{editor}{Bradley, D.} (Eds.), \bibinfo{booktitle}{Mechatronic {Futures}}. \bibinfo{publisher}{Springer International Publishing}, \bibinfo{address}{Cham}, pp. \bibinfo{pages}{59--74}.
\newblock \DOIprefix\doi{10.1007/978-3-319-32156-1_5}.
\bibitem[{Bucholtz et~al.(2020)Bucholtz, Molfino and Kolko}]{bucholtz_urbanization_2020}
\bibinfo{author}{Bucholtz, S.}, \bibinfo{author}{Molfino, E.}, \bibinfo{author}{Kolko, J.}, \bibinfo{year}{2020}.
\newblock \bibinfo{title}{The {Urbanization} {Perceptions} {Small} {Area} {Index}: {An} {Application} of {Machine} {Learning} and {Small} {Area} {Estimation} to {Household} {Survey} {Data} [working paper]}.
\newblock \URLprefix \url{https://www.huduser.gov/portal/sites/default/files/docs/UPSAI_forWeb.docx}.
\bibitem[{Caesar et~al.(2020)Caesar, Bankiti, Lang, Vora, Liong, Xu, Krishnan, Pan, Baldan and Beijbom}]{caesar_nuscenes_2020}
\bibinfo{author}{Caesar, H.}, \bibinfo{author}{Bankiti, V.}, \bibinfo{author}{Lang, A.H.}, \bibinfo{author}{Vora, S.}, \bibinfo{author}{Liong, V.E.}, \bibinfo{author}{Xu, Q.}, \bibinfo{author}{Krishnan, A.}, \bibinfo{author}{Pan, Y.}, \bibinfo{author}{Baldan, G.}, \bibinfo{author}{Beijbom, O.}, \bibinfo{year}{2020}.
\newblock \bibinfo{title}{{nuScenes}: {A} {Multimodal} {Dataset} for {Autonomous} {Driving}}, in: \bibinfo{booktitle}{2020 {IEEE}/{CVF} {Conference} on {Computer} {Vision} and {Pattern} {Recognition} ({CVPR})}, \bibinfo{publisher}{IEEE}, \bibinfo{address}{Seattle, WA, USA}. pp. \bibinfo{pages}{11618--11628}.
\newblock \DOIprefix\doi{10.1109/CVPR42600.2020.01164}.
\bibitem[{De~Gelder et~al.(2022)De~Gelder, Hof, Cator, Paardekooper, Camp, Ploeg and De~Schutter}]{de_gelder_scenario_2022}
\bibinfo{author}{De~Gelder, E.}, \bibinfo{author}{Hof, J.}, \bibinfo{author}{Cator, E.}, \bibinfo{author}{Paardekooper, J.P.}, \bibinfo{author}{Camp, O.O.D.}, \bibinfo{author}{Ploeg, J.}, \bibinfo{author}{De~Schutter, B.}, \bibinfo{year}{2022}.
\newblock \bibinfo{title}{Scenario {Parameter} {Generation} {Method} and {Scenario} {Representativeness} {Metric} for {Scenario}-{Based} {Assessment} of {Automated} {Vehicles}}.
\newblock \bibinfo{journal}{IEEE Transactions on Intelligent Transportation Systems} \bibinfo{volume}{23}, \bibinfo{pages}{18794--18807}.
\newblock \DOIprefix\doi{10.1109/TITS.2022.3154774}.
\bibitem[{{Environmental Protection Agency}()}]{environmental_protection_agency_detailed_nodate}
\bibinfo{author}{{Environmental Protection Agency}}, .
\newblock \bibinfo{title}{Detailed test information - fuel economy}.
\newblock \URLprefix \url{https://www.fueleconomy.gov/feg/fe_test_schedules.shtml}.
\bibitem[{Gloudemans et~al.(2023)Gloudemans, Wang, Gumm, Barbour and Work}]{gloudemans_interstate-24_2023}
\bibinfo{author}{Gloudemans, D.}, \bibinfo{author}{Wang, Y.}, \bibinfo{author}{Gumm, G.}, \bibinfo{author}{Barbour, W.}, \bibinfo{author}{Work, D.B.}, \bibinfo{year}{2023}.
\newblock \bibinfo{title}{The {Interstate}-24 {3D} {Dataset}: a new benchmark for {3D} multi-camera vehicle tracking} \URLprefix \url{https://arxiv.org/abs/2308.14833}, \DOIprefix\doi{10.48550/ARXIV.2308.14833}. \bibinfo{note}{publisher: arXiv Version Number: 1}.
\bibitem[{Guanetti et~al.(2018)Guanetti, Kim and Borrelli}]{guanetti_control_2018}
\bibinfo{author}{Guanetti, J.}, \bibinfo{author}{Kim, Y.}, \bibinfo{author}{Borrelli, F.}, \bibinfo{year}{2018}.
\newblock \bibinfo{title}{Control of connected and automated vehicles: {State} of the art and future challenges}.
\newblock \bibinfo{journal}{Annual Reviews in Control} \bibinfo{volume}{45}, \bibinfo{pages}{18--40}.
\bibitem[{Han et~al.(2022)Han, Karbowski and Rousseau}]{han_analytical_2022}
\bibinfo{author}{Han, J.}, \bibinfo{author}{Karbowski, D.}, \bibinfo{author}{Rousseau, A.}, \bibinfo{year}{2022}.
\newblock \bibinfo{title}{Analytical {Anticipative} {Optimal} {Drivability} {Car}-{Following} {Model}}, in: \bibinfo{booktitle}{2022 {American} {Control} {Conference} ({ACC})}, \bibinfo{publisher}{IEEE}, \bibinfo{address}{Atlanta, GA, USA}. pp. \bibinfo{pages}{4113--4118}.
\newblock \URLprefix \url{https://ieeexplore.ieee.org/document/9867588/}, \DOIprefix\doi{10.23919/ACC53348.2022.9867588}.
\bibitem[{Han et~al.(2023)Han, Shen, Jeong, Russo, Kim, Grave, Karbowski, Rousseau and Stutenberg}]{han_energy_2023}
\bibinfo{author}{Han, J.}, \bibinfo{author}{Shen, D.}, \bibinfo{author}{Jeong, J.}, \bibinfo{author}{Russo, M.D.}, \bibinfo{author}{Kim, N.}, \bibinfo{author}{Grave, J.J.}, \bibinfo{author}{Karbowski, D.}, \bibinfo{author}{Rousseau, A.}, \bibinfo{author}{Stutenberg, K.M.}, \bibinfo{year}{2023}.
\newblock \bibinfo{title}{Energy {Impact} of {Connecting} {Multiple} {Signalized} {Intersections} to {Energy}-{Efficient} {Driving}: {Simulation} and {Experimental} {Results}}.
\newblock \bibinfo{journal}{IEEE Control Systems Letters} \bibinfo{volume}{7}, \bibinfo{pages}{1297--1302}.
\newblock \DOIprefix\doi{10.1109/LCSYS.2023.3234808}.
\bibitem[{Hegde et~al.(2021)Hegde, O'Keefe, Muldoon, Gonder and Chang}]{hegde_real-world_2021}
\bibinfo{author}{Hegde, B.}, \bibinfo{author}{O'Keefe, M.}, \bibinfo{author}{Muldoon, S.}, \bibinfo{author}{Gonder, J.}, \bibinfo{author}{Chang, C.F.}, \bibinfo{year}{2021}.
\newblock \bibinfo{title}{Real-{World} {Driving} {Features} for {Identifying} {Intelligent} {Driver} {Model} {Parameters}}, pp. \bibinfo{pages}{2021--01--0436}.
\newblock \URLprefix \url{https://www.sae.org/content/2021-01-0436/}, \DOIprefix\doi{10.4271/2021-01-0436}.
\bibitem[{Houston et~al.(2020)Houston, Zuidhof, Bergamini, Ye, Chen, Jain, Omari, Iglovikov and Ondruska}]{houston_one_2020}
\bibinfo{author}{Houston, J.}, \bibinfo{author}{Zuidhof, G.}, \bibinfo{author}{Bergamini, L.}, \bibinfo{author}{Ye, Y.}, \bibinfo{author}{Chen, L.}, \bibinfo{author}{Jain, A.}, \bibinfo{author}{Omari, S.}, \bibinfo{author}{Iglovikov, V.}, \bibinfo{author}{Ondruska, P.}, \bibinfo{year}{2020}.
\newblock \bibinfo{title}{One {Thousand} and {One} {Hours}: {Self}-driving {Motion} {Prediction} {Dataset}} \URLprefix \url{https://arxiv.org/abs/2006.14480}, \DOIprefix\doi{10.48550/ARXIV.2006.14480}. \bibinfo{note}{publisher: arXiv Version Number: 2}.
\bibitem[{Jeong et~al.(2023)Jeong, Dudekula, Kandaswamy, Karbowski, Han and Naber}]{jeong_-track_2023}
\bibinfo{author}{Jeong, J.}, \bibinfo{author}{Dudekula, A.B.}, \bibinfo{author}{Kandaswamy, E.}, \bibinfo{author}{Karbowski, D.}, \bibinfo{author}{Han, J.}, \bibinfo{author}{Naber, J.}, \bibinfo{year}{2023}.
\newblock \bibinfo{title}{On-{Track} {Demonstration} of {Automated} {Eco}-{Driving} {Control} for an {Electric} {Vehicle}}, \bibinfo{address}{Detroit, Michigan, United States}. pp. \bibinfo{pages}{2023--01--0221}.
\newblock \DOIprefix\doi{10.4271/2023-01-0221}.
\bibitem[{Kesting and Treiber(2008)}]{kesting_calibrating_2008}
\bibinfo{author}{Kesting, A.}, \bibinfo{author}{Treiber, M.}, \bibinfo{year}{2008}.
\newblock \bibinfo{title}{Calibrating {Car}-{Following} {Models} by {Using} {Trajectory} {Data}: {Methodological} {Study}}.
\newblock \bibinfo{journal}{Transportation Research Record: Journal of the Transportation Research Board} \bibinfo{volume}{2088}, \bibinfo{pages}{148--156}.
\newblock \DOIprefix\doi{10.3141/2088-16}.
\bibitem[{Krajewski et~al.(2018)Krajewski, Bock, Kloeker and Eckstein}]{krajewski_highd_2018}
\bibinfo{author}{Krajewski, R.}, \bibinfo{author}{Bock, J.}, \bibinfo{author}{Kloeker, L.}, \bibinfo{author}{Eckstein, L.}, \bibinfo{year}{2018}.
\newblock \bibinfo{title}{The {highD} {Dataset}: {A} {Drone} {Dataset} of {Naturalistic} {Vehicle} {Trajectories} on {German} {Highways} for {Validation} of {Highly} {Automated} {Driving} {Systems}}, in: \bibinfo{booktitle}{2018 21st {International} {Conference} on {Intelligent} {Transportation} {Systems} ({ITSC})}, \bibinfo{publisher}{IEEE}, \bibinfo{address}{Maui, HI}. pp. \bibinfo{pages}{2118--2125}.
\newblock \DOIprefix\doi{10.1109/ITSC.2018.8569552}.
\bibitem[{Liu and Zhang(2022)}]{liu_learning_2022}
\bibinfo{author}{Liu, C.}, \bibinfo{author}{Zhang, W.}, \bibinfo{year}{2022}.
\newblock \bibinfo{title}{Learning the {Driver} {Acceleration}/{Deceleration} {Behavior} {Under} {High}-{Speed} {Environments} {From} {Naturalistic} {Driving} {Data}}.
\newblock \bibinfo{journal}{IEEE Intelligent Transportation Systems Magazine} \bibinfo{volume}{14}, \bibinfo{pages}{78--91}.
\newblock \DOIprefix\doi{10.1109/MITS.2020.3014115}.
\bibitem[{Menneni et~al.(2008)Menneni, Sun and Vortisch}]{menneni_microsimulation_2008}
\bibinfo{author}{Menneni, S.}, \bibinfo{author}{Sun, C.}, \bibinfo{author}{Vortisch, P.}, \bibinfo{year}{2008}.
\newblock \bibinfo{title}{Microsimulation {Calibration} {Using} {Speed}-{Flow} {Relationships}}.
\newblock \bibinfo{journal}{Transportation Research Record: Journal of the Transportation Research Board} \bibinfo{volume}{2088}, \bibinfo{pages}{1--9}.
\newblock \DOIprefix\doi{10.3141/2088-01}.
\bibitem[{Mersky and Samaras(2016)}]{mersky_fuel_2016}
\bibinfo{author}{Mersky, A.C.}, \bibinfo{author}{Samaras, C.}, \bibinfo{year}{2016}.
\newblock \bibinfo{title}{Fuel economy testing of autonomous vehicles}.
\newblock \bibinfo{journal}{Transportation Research Part C: Emerging Technologies} \bibinfo{volume}{65}, \bibinfo{pages}{31--48}.
\newblock \DOIprefix\doi{10.1016/j.trc.2016.01.001}.
\bibitem[{Negash and Yang(2023)}]{negash_driver_2023}
\bibinfo{author}{Negash, N.M.}, \bibinfo{author}{Yang, J.}, \bibinfo{year}{2023}.
\newblock \bibinfo{title}{Driver {Behavior} {Modeling} {Toward} {Autonomous} {Vehicles}: {Comprehensive} {Review}}.
\newblock \bibinfo{journal}{IEEE Access} \bibinfo{volume}{11}, \bibinfo{pages}{22788--22821}.
\newblock \DOIprefix\doi{10.1109/ACCESS.2023.3249144}.
\bibitem[{Riedmaier et~al.(2020)Riedmaier, Ponn, Ludwig, Schick and Diermeyer}]{riedmaier_survey_2020}
\bibinfo{author}{Riedmaier, S.}, \bibinfo{author}{Ponn, T.}, \bibinfo{author}{Ludwig, D.}, \bibinfo{author}{Schick, B.}, \bibinfo{author}{Diermeyer, F.}, \bibinfo{year}{2020}.
\newblock \bibinfo{title}{Survey on {Scenario}-{Based} {Safety} {Assessment} of {Automated} {Vehicles}}.
\newblock \bibinfo{journal}{IEEE Access} \bibinfo{volume}{8}, \bibinfo{pages}{87456--87477}.
\newblock \DOIprefix\doi{10.1109/ACCESS.2020.2993730}.
\bibitem[{{Safety Pilot Model Deployment}(2014)}]{safety_pilot_model_deployment_safety_2014}
\bibinfo{author}{{Safety Pilot Model Deployment}}, \bibinfo{year}{2014}.
\newblock \bibinfo{title}{Safety {Pilot} {Model} {Deployment} {Data}}.
\newblock \URLprefix \url{https://data.transportation.gov/d/a7qq-9vfe}, \DOIprefix\doi{10.21949/1504482}.
\bibitem[{Scanlon et~al.(2018)Scanlon, Sherony and Gabler}]{scanlon_models_2018}
\bibinfo{author}{Scanlon, J.M.}, \bibinfo{author}{Sherony, R.}, \bibinfo{author}{Gabler, H.C.}, \bibinfo{year}{2018}.
\newblock \bibinfo{title}{Models of {Driver} {Acceleration} {Behavior} {Prior} to {Real}-{World} {Intersection} {Crashes}}.
\newblock \bibinfo{journal}{IEEE Transactions on Intelligent Transportation Systems} \bibinfo{volume}{19}, \bibinfo{pages}{774--786}.
\newblock \DOIprefix\doi{10.1109/TITS.2017.2699079}.
\bibitem[{Schwarz and Wang(2022)}]{schwarz_role_2022}
\bibinfo{author}{Schwarz, C.}, \bibinfo{author}{Wang, Z.}, \bibinfo{year}{2022}.
\newblock \bibinfo{title}{The {Role} of {Digital} {Twins} in {Connected} and {Automated} {Vehicles}}.
\newblock \bibinfo{journal}{IEEE Intelligent Transportation Systems Magazine} \bibinfo{volume}{14}, \bibinfo{pages}{41--51}.
\newblock \DOIprefix\doi{10.1109/MITS.2021.3129524}.
\bibitem[{Sun et~al.(2020)Sun, Kretzschmar, Dotiwalla, Chouard, Patnaik, Tsui, Guo, Zhou, Chai, Caine, Vasudevan, Han, Ngiam, Zhao, Timofeev, Ettinger, Krivokon, Gao, Joshi, Zhang, Shlens, Chen and Anguelov}]{sun_scalability_2020}
\bibinfo{author}{Sun, P.}, \bibinfo{author}{Kretzschmar, H.}, \bibinfo{author}{Dotiwalla, X.}, \bibinfo{author}{Chouard, A.}, \bibinfo{author}{Patnaik, V.}, \bibinfo{author}{Tsui, P.}, \bibinfo{author}{Guo, J.}, \bibinfo{author}{Zhou, Y.}, \bibinfo{author}{Chai, Y.}, \bibinfo{author}{Caine, B.}, \bibinfo{author}{Vasudevan, V.}, \bibinfo{author}{Han, W.}, \bibinfo{author}{Ngiam, J.}, \bibinfo{author}{Zhao, H.}, \bibinfo{author}{Timofeev, A.}, \bibinfo{author}{Ettinger, S.}, \bibinfo{author}{Krivokon, M.}, \bibinfo{author}{Gao, A.}, \bibinfo{author}{Joshi, A.}, \bibinfo{author}{Zhang, Y.}, \bibinfo{author}{Shlens, J.}, \bibinfo{author}{Chen, Z.}, \bibinfo{author}{Anguelov, D.}, \bibinfo{year}{2020}.
\newblock \bibinfo{title}{Scalability in {Perception} for {Autonomous} {Driving}: {Waymo} {Open} {Dataset}}, in: \bibinfo{booktitle}{2020 {IEEE}/{CVF} {Conference} on {Computer} {Vision} and {Pattern} {Recognition} ({CVPR})}, \bibinfo{publisher}{IEEE}, \bibinfo{address}{Seattle, WA, USA}. pp. \bibinfo{pages}{2443--2451}.
\newblock \DOIprefix\doi{10.1109/CVPR42600.2020.00252}.
\bibitem[{Toledo(2007)}]{toledo_driving_2007}
\bibinfo{author}{Toledo, T.}, \bibinfo{year}{2007}.
\newblock \bibinfo{title}{Driving {Behaviour}: {Models} and {Challenges}}.
\newblock \bibinfo{journal}{Transport Reviews} \bibinfo{volume}{27}, \bibinfo{pages}{65--84}.
\newblock \DOIprefix\doi{10.1080/01441640600823940}.
\bibitem[{{U.S. Department Of Transportation Federal Highway Administration}(2017)}]{us_department_of_transportation_federal_highway_administration_next_2017}
\bibinfo{author}{{U.S. Department Of Transportation Federal Highway Administration}}, \bibinfo{year}{2017}.
\newblock \bibinfo{title}{Next {Generation} {Simulation} ({NGSIM}) {Vehicle} {Trajectories} and {Supporting} {Data}}.
\newblock \URLprefix \url{https://data.transportation.gov/d/8ect-6jqj}, \DOIprefix\doi{10.21949/1504477}.
\bibitem[{Vahidi and Sciarretta(2018)}]{vahidi_energy_2018}
\bibinfo{author}{Vahidi, A.}, \bibinfo{author}{Sciarretta, A.}, \bibinfo{year}{2018}.
\newblock \bibinfo{title}{Energy saving potentials of connected and automated vehicles}.
\newblock \bibinfo{journal}{Transportation Research Part C: Emerging Technologies} \bibinfo{volume}{95}, \bibinfo{pages}{822--843}.
\bibitem[{Wang et~al.(2019)Wang, Yang and Hurwitz}]{wang_analysis_2019}
\bibinfo{author}{Wang, X.}, \bibinfo{author}{Yang, M.}, \bibinfo{author}{Hurwitz, D.}, \bibinfo{year}{2019}.
\newblock \bibinfo{title}{Analysis of cut-in behavior based on naturalistic driving data}.
\newblock \bibinfo{journal}{Accident Analysis \& Prevention} \bibinfo{volume}{124}, \bibinfo{pages}{127--137}.
\newblock \DOIprefix\doi{10.1016/j.aap.2019.01.006}.
\bibitem[{Yan et~al.(2023)Yan, Zou, Feng, Zhu, Sun and Liu}]{yan_learning_2023}
\bibinfo{author}{Yan, X.}, \bibinfo{author}{Zou, Z.}, \bibinfo{author}{Feng, S.}, \bibinfo{author}{Zhu, H.}, \bibinfo{author}{Sun, H.}, \bibinfo{author}{Liu, H.X.}, \bibinfo{year}{2023}.
\newblock \bibinfo{title}{Learning naturalistic driving environment with statistical realism}.
\newblock \bibinfo{journal}{Nature Communications} \bibinfo{volume}{14}, \bibinfo{pages}{2037}.
\newblock \DOIprefix\doi{10.1038/s41467-023-37677-5}.
\bibitem[{Zhao et~al.(2017)Zhao, Lam, Peng, Bao, LeBlanc, Nobukawa and Pan}]{zhao_accelerated_2017}
\bibinfo{author}{Zhao, D.}, \bibinfo{author}{Lam, H.}, \bibinfo{author}{Peng, H.}, \bibinfo{author}{Bao, S.}, \bibinfo{author}{LeBlanc, D.J.}, \bibinfo{author}{Nobukawa, K.}, \bibinfo{author}{Pan, C.S.}, \bibinfo{year}{2017}.
\newblock \bibinfo{title}{Accelerated {Evaluation} of {Automated} {Vehicles} {Safety} in {Lane}-{Change} {Scenarios} {Based} on {Importance} {Sampling} {Techniques}}.
\newblock \bibinfo{journal}{IEEE Transactions on Intelligent Transportation Systems} \bibinfo{volume}{18}, \bibinfo{pages}{595--607}.
\newblock \DOIprefix\doi{10.1109/TITS.2016.2582208}.

\end{thebibliography}



\end{document}